\documentclass[letterpaper,onecolumn,pra,
aps,showpacs,groupedaddress,floatfix,notitlepage]{revtex4-2}
\usepackage{microtype}
\usepackage[latin1]{inputenc}
\usepackage{bm}
\usepackage[usenames]{color}
\usepackage[dvipsnames]{xcolor}
\usepackage{multirow}
\usepackage{framed}
\usepackage{amssymb}
\usepackage{amsbsy}
\usepackage{mathtools}
\usepackage{amsmath}
\usepackage{stmaryrd}
\usepackage{graphicx}
\usepackage{epsfig}
\usepackage{placeins}
\usepackage{bbold}
\usepackage{braket}
\usepackage{blindtext}
\usepackage[colorlinks,linkcolor=blue,citecolor=blue,urlcolor=blue]{hyperref}
\usepackage{tcolorbox}
\usepackage{xspace}
\usepackage[capitalize]{cleveref}
\usepackage{dsfont}
\usepackage[caption=false]{subfig}
\usepackage[version=4]{mhchem}
\usepackage[parfill]{parskip}
\usepackage{tikz}
\usetikzlibrary{math}
\usetikzlibrary{arrows.meta}
\usepackage{wrapfig}
\usepackage{tabularx}
\usepackage{booktabs}
\usepackage{floatrow}

\makeatletter

\newcommand{\tr}[1]{\ensuremath{\mathrm{tr}\left[#1\right]}\xspace}
\newcommand{\renyi}{R\'{e}nyi\xspace}
\newcommand{\diff}{\ensuremath{\mathrm{d}}\xspace}

\newsavebox{\measurebox}

\begin{document}
\title{Quantum simulation using noisy unitary circuits and measurements}

\author{Oliver Lunt}
\affiliation{School of Physics and Astronomy, University of Birmingham, Birmingham, B15 2TT, UK}

\author{Jonas Richter}
\affiliation{Department of Physics and Astronomy, University College London, 
Gower Street, London, WC1E 6BT, UK}

\author{Arijeet Pal}
\affiliation{Department of Physics and Astronomy, University College London, 
Gower Street, London, WC1E 6BT, UK}

\date{\today}

\begin{abstract}
Many-body quantum systems are notoriously hard to study theoretically due to the exponential growth of their Hilbert space. It is also challenging to probe the quantum correlations in many-body states in experiments due to their sensitivity to external noise. Using synthetic quantum matter to simulate quantum systems has opened new ways of probing quantum many-body systems with unprecedented control, and of engineering phases of matter which are otherwise hard to find in nature. Noisy quantum circuits have become an important cornerstone of our understanding 
of quantum many-body dynamics. In particular, random circuits act as minimally 
structured toy models for chaotic nonintegrable quantum systems, 
faithfully reproducing some of their universal properties. Crucially, in 
contrast to the full microscopic model, random circuits can be analytically tractable under a reasonable set of assumptions, thereby providing invaluable insights into questions which 
might be out of reach even for state-of-the-art numerical techniques. Here, we 
give an overview of two classes of dynamics studied using random-circuit models, with a particular focus on 
the dynamics of quantum entanglement. We will especially pay attention to  
potential near-term applications of random-circuit models on noisy-intermediate 
scale quantum (NISQ) devices. In this context, we cover hybrid circuits 
consisting of unitary gates interspersed with nonunitary projective 
measurements, hosting an entanglement phase transition from a volume-law to an area-law phase 
of the steady-state entanglement. Moreover, we consider random-circuit sampling 
experiments and discuss the usefulness of random quantum states 
for simulating quantum many-body dynamics on NISQ devices by leveraging the 
concept of quantum typicality. We highlight how emergent hydrodynamics can be studied by utilizing random quantum states generated by chaotic circuits.

\end{abstract}

\maketitle

\section{Introduction}\label{Sec::Intro}

Recent developments in quantum many-body physics have highlighted the importance of entanglement in understanding phases of matter in and out of equilibrium \cite{Dalessio2016, NandkishoreHuse_review, Abanin_MBL}. Even limited control of entanglement in quantum circuits has provided an opportunity for the fusion of ideas from quantum information science and condensed matter physics, and an exciting platform for testing our knowledge of quantum many-body physics \cite{Preskill2018, potter2021entanglement}. This has been particularly constructive for understanding phenomena far away from the ground state of quantum many-body systems such as quantum chaos \cite{nahumQuantumEntanglementGrowth2017, vonkeyserlingkOperatorHydrodynamicsOTOCs2018, Khemani_Hydro_PRX2018, Bertini_PRL2018, Chan_PRX2018} and many-body localization \cite{gornyi2005interacting, basko2006metal, Pal_MBL2010, Pollmann_MBLTN_2016, Wahl_PRX2017, wahl2019signatures}. The quantum simulation of complex quantum phenomena exploring the exponentially large Hilbert space have been enriched by using entanglement as a probe. Entanglement provides a non-trivial parametrization of the many-body states which has ramifications for their simulability \cite{nappEfficientClassicalSimulation2019, noh2020efficient, azad2021phase}. For non-equilibrium phenomena the dynamics of entanglement can exhibit universal features characterizing the phase of matter and its coarse grained properties, which can be a useful tool for visualizing macroscopic quantum coherence.

Realizations of unitary quantum gates and fast measurements achieved in several physical systems, such as superconducting circuit QED systems \cite{murch2013observing, roch2014observation, minev2019catch} and trapped ions \cite{Sauter_1986, Bergquist_1986, noel2021observation}, have opened new avenues for quantum simulation. The low error rates in these unitary gates allow coherent evolution of many-body states over relatively long time scales enabling the study of quantum correlations in the form of entanglement. This experimental effort is in its early stages with $10$s of qubits, but nonetheless is already in the realm of providing an advantage for simulating quantum systems compared to classical supercomputers. The quantum control of the microscopic degrees of freedom has also opened the avenue to realize exotic entangled states stabilized by non-equilibrium effects \cite{lavasaniMeasurementinducedTopologicalEntanglement2021, sangMeasurementProtectedQuantum2020, Claeys_DUC2021, Ippoliti_NISQ2021}. 
Digital quantum simulation using a gate-based architecture complements the earlier achievements in analogue quantum simulation using ultra-cold atoms \cite{bloch2012quantum, Bloch_RMP2008, greiner2002quantum, schreiber2015observation, simon2011quantum, kaufman2016quantum, bernien2017probing} due to its connections to quantum error correction and complexity and provides a new lens for classifying quantum phases of matter \cite{choiQuantumErrorCorrection2020, liStatisticalMechanicsQuantum2020, fanSelfOrganizedErrorCorrection2020, gullansQuantumCodingLowdepth2020}. 

The study of quantum circuits has enriched our understanding of quantum chaos, providing a toy model for describing entanglement dynamics in these systems \cite{Nahum_PRX2018, vonkeyserlingkOperatorHydrodynamicsOTOCs2018}. In contrast to the eigenstate thermalization hypothesis which is applicable to Hamiltonian and Floquet systems \cite{Deutsch_PRA91, Srednicki_PRE94, Dalessio2016}, where one is often limited to exact diagonalization to evaluate eigenstates, unitary circuits comprising of randomly chosen gates offer a simplified effective description for chaos. For a certain class of circuits, known as dual-unitary circuits \cite{Bertini_PRL2019, Gopalakrishnan_PRB2019, Piroli_PRB2020, Claeys_DUC2021}, constrained to be unitary along dual directions of space and time, the dynamics in the circuit are further simplified for certain observables while retaining their quantum chaotic properties. On introducing global conservation laws, the random unitary prescription naturally lends itself to describing the emergent hydrodynamics of the conserved quantities opening a new platform for studying quantum hydrodynamics.

These circuits are also fertile playground for studying dynamics induced by measurements on the qubits \cite{liQuantumZenoEffect2018, skinnerMeasurementInducedPhaseTransitions2019, chanUnitaryprojectiveEntanglementDynamics2019}. It opens a new regime for probing dynamics of entanglement of open quantum systems far from equilibrium. Despite measurements, the quantum trajectories of a many-body system for weak measurements can become highly entangled and far from a classical product state. In the strong measurement regime the trajectories become unentangled and lose their quantum correlations. Many questions related to the nature of the phase transitions between these two phases continue to be actively investigated \cite{liMeasurementdrivenEntanglementTransition2019, szyniszewskiEntanglementTransitionVariablestrength2019, chenEmergentConformalSymmetry2020, zabaloCriticalPropertiesMeasurementinduced2020, ippolitiEntanglementPhaseTransitions2021, luntMeasurementinducedCriticalityEntanglement2021, Alberton_PRL2021, zabalo2021operator, Sang_PRXQ2021, Tang_PRB2021}.  The entanglement preserving phase can have error correcting properties and is being considered for stabilizing novel many-body states with multipartite entanglement by monitoring local degrees of freedom \cite{gullansDynamicalPurificationPhase2020, hastingsDynamicallyGeneratedLogical2021a, verresen2021efficiently}.

\begin{figure}[tb]
 \centering
 \includegraphics[width=0.8\textwidth]{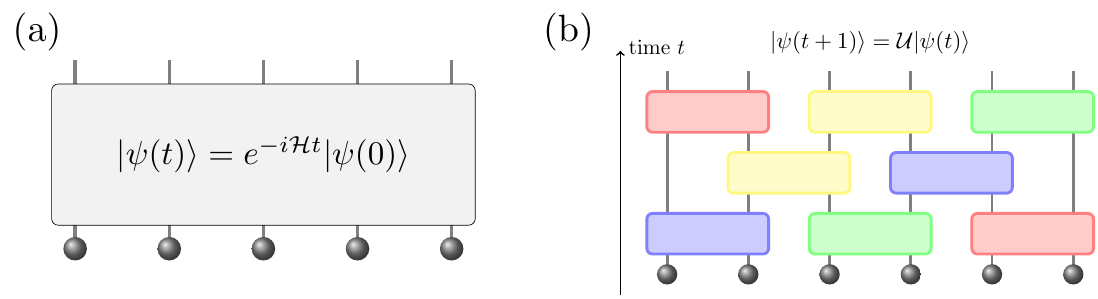}
 \caption{\textbf{(a)} Given a (chaotic) quantum many-body system described by a Hamiltonian ${\cal H}$, the continuous time evolution of a quantum state $\ket{\psi}$ is generated by the unitary operator $\exp(-i{\cal H}t)$. \textbf{(b)} Random circuits act as minimal models to describe the properties of chaotic quantum many-body systems, with local gates acting on the degrees of freedom, with the time evolution now being discrete. 
 }
 \label{Fig_Circuit_1}
\end{figure}

In this chapter, we 
give an overview of random-circuit models, with a particular focus on 
the dynamics of quantum entanglement. We will especially pay attention to  
potential near-term applications of random-circuit models on noisy-intermediate 
scale quantum (NISQ) devices. To this end, consider a quantum many-body systems with 
physical degrees of freedom defined on discrete lattice sites. The degrees of 
freedom can for instance be quantum spins, in which case the local 
Hilbert-space dimension is $d = 2s+1$ with $s$ being the spin quantum number, 
but also fermionic or bosonic particles. In case of an isolated quantum system,  
its time evolution is unitary and governed by the time-dependent Schr\"odinger 
equation. Specifically, given some out-of-equilibrium initial state 
$\ket{\psi(0)}$, the time-evolved state at a later time $t$ follows as, cf.\ 
Fig.\ \ref{Fig_Circuit_1}~(a),   
\begin{equation}\label{Eq::TimeEvo}
 \ket{\psi(t)} = e^{-i{\cal H}t} \ket{\psi(0)}\ , 
\end{equation}
where ${\cal H}$ denotes the Hamiltonian of the system. In practice, however, the evaluation of Eq.\ 
\eqref{Eq::TimeEvo} is challenging due to the exponentially growing Hilbert space with increasing system size 
$L$ ($D = d^L = 2^L$ for quantum spins with $s = 1/2$, i.e., 
qubits). Even though significant progress has been achieved in solving Eq.\ 
\eqref{Eq::TimeEvo} also for large quantum systems due to the development of 
sophisticated numerical methods, especially matrix-product state techniques (see e.g.\ \cite{Paeckel_2019}), 
such approaches are typically limited by the growth of quantum entanglement 
during the time evolution,   
\begin{equation}
S(t) = -\text{tr}[\rho_A \ln \rho_A]\ ,\qquad \rho_A = 
\text{Tr}_B(\ket{\psi(t)}\bra{\psi(t)})\ , 
\end{equation}
where $\rho_A$ denotes the reduced density matrix for a bipartition of the 
system into subsystems $A$ and $B$. In particular, $S(t)$ is typically expected 
to increase linearly in time, $S(t) \propto t$ \cite{Kim_2013}, which causes the computational 
requirements to faithfully describe the state $\ket{\psi(t)}$ to grow 
exponentially. Note, however, that there are also cases where $S(t)$ builds up 
slower such as disordered models exhibiting many-body localization \cite{Abanin_MBL}. 

While the out-of-equilibrium dynamics of quantum many-body systems 
are hard to analyze, random circuits provide
minimally structured models to describe their properties. The 
circuits consist of unitary gates acting locally on the degrees of freedom, 
for instance in a brickwall pattern akin to well-known Trotter decompositions, 
see Fig.\ \ref{Fig_Circuit_1}~(b). The gates are not chosen to 
reproduce the properties of a particular Hamiltonian, but rather to capture the 
universal aspects of quantum chaotic dynamics. In particular, the gates 
are drawn at random, for instance the two-site gates in Fig.\ 
\ref{Fig_Circuit_1}~(b) are $d^2 \times d^2$ matrices, which could be drawn 
from the unitarily-invariant Haar measure or from the Clifford group \cite{nahumQuantumEntanglementGrowth2017}. Moreover, 
it is possible to consider the impact of symmetries, such as conservation of total 
magnetization or dipole moment, as well as kinetic constraints, by suitably 
choosing the gates (e.g., Haar-random matrices that are block-diagonal) \cite{vonkeyserlingkOperatorHydrodynamicsOTOCs2018, Khemani_Hydro_PRX2018, Pai_2019}. In 
contrast to the intrinsically continuous time evolution of quantum systems 
[cf.\ Eq.\ \eqref{Eq::TimeEvo}], the time evolution in circuit models is  
discrete, 
\begin{equation}
 \ket{\psi(t+1)} = {\cal U}\ket{\psi(t)}\ , 
\end{equation}
where ${\cal U}$ is the full unitary operator of one layer of the circuit. 
Nevertheless, analogous to the Hamiltonian time evolution, the entanglement 
entropy $S(t)$ is found to grow very rapidly in random-circuit models \cite{nahumQuantumEntanglementGrowth2017,vonkeyserlingkOperatorHydrodynamicsOTOCs2018}.
Crucially, random circuits not only capture the essential features of 
chaotic quantum many-body dynamics, but in some cases also allow for analytical 
solutions of the dynamics. The latter makes these models very valuable to gain 
important insights that are out of reach for currently available numerical 
approaches. 

In this chapter, we especially focus on two particular kinds of circuit
models: In Sec.\ \ref{sec::Measurement}, we consider monitored circuits, where 
unitary gates are interspersed with local projective measurements. These 
circuits can be understood as toy models to describe the interaction of quantum systems with their environment, 
which constantly ``measures'' the system leading to decoherence. In particular, 
it has been found that the competition between unitary gates and projective 
measurements leads to a dynamical phase transition between a volume-law phase 
of $S(t)$ and an area-law phase of $S(t)$. Furthermore, in Sec.\ 
\ref{Sec::NISQ}, we consider pseudo-random circuits consisting of one- and 
two-qubit gates drawn from a set of elementary gates available on today's NISQ devices. These circuits have recently 
gained importance to demonstrate a quantum computational advantage, i.e., to 
perform a computational task with a quantum computer that is out of reach for
classical machines. We will discuss the main ideas of the random-circuit 
sampling task that was implemented, and also describe how such circuits can 
be useful in the context of simulations of quantum many-body dynamics on 
NISQ devices.
 
\section{Measurement-induced entanglement transitions in hybrid quantum circuits}
\label{sec::Measurement}

In a many-body quantum system, the entanglement entropy typically follows a \textit{scaling law}, where the entropy $S(A)$ of a contiguous subsystem $A$ scales with some geometric property of $A$. There are several examples of scaling laws. A \textit{volume law} means the entropy $S(A) \propto |A| +  \cdots$ scales to leading order with the size of $A$, whereas an \textit{area law} means the entropy $S(A) \propto |\partial A| + \cdots$ scales with the size of the boundary of $A$. The highly entangled volume law states can be found, for example, as the steady states of chaotic quantum dynamics~\cite{Dalessio2016}, while an area law is often found in states with fast decay of correlations~\cite{brandaoAreaLawEntanglement2013}, such as the ground states of gapped Hamiltonians~\cite{hastingsAreaLawOnedimensional2007}. There are also scaling laws intermediate between volume and area laws; for example, the entropy of the ground state of a 1+1D conformal field theory scales \textit{logarithmically} as $S(A) \propto \log{|A|} + \cdots$ \cite{calabreseEntanglementEntropyQuantum2004}.

An \textit{entanglement transition} is then a phase transition in which the scaling behavior of the entanglement entropy changes in some state of interest, such as an energy eigenstate or the steady state of some quantum dynamics. In this section we will mainly be interested in a particular class of entanglement transitions, known as \textit{measurement-induced} transitions~\cite{liQuantumZenoEffect2018,chanUnitaryprojectiveEntanglementDynamics2019,skinnerMeasurementInducedPhaseTransitions2019,szyniszewskiEntanglementTransitionVariablestrength2019,liMeasurementdrivenEntanglementTransition2019,nappEfficientClassicalSimulation2019,zabaloCriticalPropertiesMeasurementinduced2020,fanSelfOrganizedErrorCorrection2020,gullansDynamicalPurificationPhase2020,baoTheoryPhaseTransition2020,jianMeasurementinducedCriticalityRandom2020,liConformalInvarianceQuantum2021,shtankoClassicalModelsEntanglement2020,lavasaniMeasurementinducedTopologicalEntanglement2021,sangMeasurementProtectedQuantum2020,szyniszewskiUniversalityEntanglementTransitions2020,zhangNonuniversalEntanglementLevel2020,choiQuantumErrorCorrection2020,turkeshiMeasurementinducedCriticalityDimensional2020,gullansScalableProbesMeasurementInduced2020,nahumMeasurementEntanglementPhase2021,caoEntanglementFermionChain2019,tangMeasurementinducedPhaseTransition2020,gotoMeasurementInducedTransitionsEntanglement2020,Alberton_PRL2021,luntMeasurementinducedEntanglementTransitions2020,langEntanglementTransitionProjective2020,chenEmergentConformalSymmetry2020,liuNonunitaryDynamicsSachdevYeKitaev2020,fujiMeasurementinducedQuantumCriticality2020,ippolitiEntanglementPhaseTransitions2021,vanregemortelEntanglementEntropyScaling2021,aharonovQuantumClassicalPhase2000,vijayMeasurementDrivenPhaseTransition2020,nahumEntanglementDynamicsDiffusionannihilation2020,liStatisticalMechanicsQuantum2020,rossiniMeasurementinducedDynamicsManybody2020,gullansScalableProbesMeasurementInduced2020,gullansQuantumCodingLowdepth2020,fidkowskiHowDynamicalQuantum2021,maimbourgBathinducedZenoLocalization2020,iaconisMeasurementinducedPhaseTransitions2020,ippolitiPostselectionFreeEntanglementDynamics2021,lavasaniTopologicalOrderCriticality2020,Sang_PRXQ2021,shiEntanglementNegativityCritical2020,gopalakrishnanEntanglementPurificationTransitions2021,luntMeasurementinducedCriticalityEntanglement2021,liStatisticalMechanicsModel2021,hanMeasurementInducedCriticality2021,chenNonunitaryFreeBoson2021,iaconisMultifractalityNonunitaryRandom2021,sharmaMeasurementinducedCriticalityExtended2021,sierantDissipativeFloquetDynamics2021,sierantUniversalBehaviorMultifractality2021,turkeshiEntanglementTransitionsStochastic2021,turkeshiMeasurementinducedCriticalityDatastructure2021,turkeshiMeasurementinducedCriticalityDimensional2020,turkeshiMeasurementinducedEntanglementTransitions2021}. These occur in the steady state of \textit{nonunitary} dynamics, where the nonunitarity is a result of quantum measurements being applied to the system at a constant rate. This setup can emerge quite naturally as a model for the interaction between a system and its environment. When the overall dynamics includes a mixture of measurements and also unitary dynamics, such as Hamiltonian time evolution representing a system's internal dynamics, we will refer to this as `hybrid quantum dynamics'. Unitary dynamics often provides a natural route to generating entanglement, which is typically (though not always!) destroyed by measurements. However, it is known that entanglement transitions can also occur in \textit{measurement-only} models~\cite{lavasaniMeasurementinducedTopologicalEntanglement2021,ippolitiEntanglementPhaseTransitions2021,lavasaniTopologicalOrderCriticality2020,vanregemortelEntanglementEntropyScaling2021,langEntanglementTransitionProjective2020}, where the entanglement is generated as a result of frustration between the measurements.

\subsection{Quantum trajectories}

When modeling the effect of an external environment on a quantum system, there are several complementary approaches. In the `mixed state approach', we take into account the fact that some information about the system will leak into the environment by modelling the state of the system with a density matrix. The mixed state $\rho(t)$ will then evolve in time according to some master equation. An alternative viewpoint is the `quantum trajectories approach', where we model the system with a pure state, and treat the effect of the environment through repeated measurements of the state. These approaches are equivalent in the sense that one can map between the two pictures: averaging over the possible quantum trajectories recovers the master equation for the density matrix $\rho$, and one can `unravel' the master equation to focus on individual quantum trajectories. Note that this unraveling is in general not unique; for example, consider the following trajectory equations, both describing Hamiltonian evolution of a fermionic chain undergoing continuous weak measurements~\cite{Alberton_PRL2021}. The first class of trajectories, known as quantum state diffusion, is described by
\begin{equation}
	\mathrm{d}\!\ket{\psi} = \left[-i H \diff t + \sum_{l} \left(\dfrac{\gamma}{2} \hat{M}_{l,t}^{2} \diff t + \xi_{l,t} \hat{M}_{l,t} \right) \right] \ket{\psi},
\end{equation}
where $\xi_{l,t}$ is a real-valued Gaussian variable with zero mean and covariance $\overline{\xi_{l,t} \xi_{m,t^{'}}} = \gamma \diff t \delta_{l,m} \delta(t-t^{'})$, and $\hat{M}_{l,t} = n_{l} - \langle n_{l} \rangle_{t}$ with $n_{l}$ the local fermion occupation number. We could also consider the class of `quantum jump' trajectories, given by
\begin{equation}
	\mathrm{d}\!\ket{\psi} = \left[-i H \diff t + \sum_{l} \xi_{l,t} \left( \dfrac{n_{l}}{\sqrt{\langle n_{l} \rangle_{t}}} - 1 \right) \right] \ket{\psi}
\end{equation}
for a state with conserved particle number, where now the noise is defined by $\xi_{l,t}^{2} = \xi_{l,t}$ and $\overline{\xi_{l,t}} = \gamma \diff t \langle n_{l} \rangle_{t}$. Upon averaging the state $\ket{\psi} \bra{\psi}$ over the noise $\{\xi_{l,t}\}$ to get the density matrix $\rho = \overline{\ket{\psi}\bra{\psi}}$, both of these models give rise to the same master equation
\begin{equation}
	\partial_{t} \rho = -i[H,\rho] + \gamma \sum_{l}\left( 2 n_{l} \rho n_{l} - \{n_{l}^{2},\rho\}\right).
\end{equation}
However, that is not to say that these two forms of trajectory are completely equivalent. In particular, some quantities do not commute with the average over states, in particular if they are a \textit{nonlinear} function of the state --- in other words, we can have functions $f$ of the state with
\begin{equation}
	\underset{i}{\mathbb{E}}\left[f\left(\ket{\psi_{i}}\bra{\psi_{i}}\right)\right] \neq f\left( \underset{i}{\mathbb{E}}\left[ \ket{\psi_{i}}\bra{\psi_{i}}\right]\right).
\end{equation}
For our purposes the main example of such a quantity will be the entanglement entropy $S(A) = -\tr{\rho_{A} \ln{\rho_{A}}}$, but the same is also true for connected correlation functions of observables, $\langle O_{A} O_{B} \rangle_{c} \equiv \langle O_{A} O_{B} \rangle - \langle O_{A} \rangle \langle O_{B} \rangle$. One surprising aspect of measurement-induced transitions is that there can be phase transitions in quantities calculated along individual quantum trajectories that are completely invisible if one only looks at linear functions of the state. In particular, this means that for these phase transitions there can be no order parameter $\langle O \rangle = \tr{O \rho}$ given by an expectation value taken with the average state. In the rest of this section, we will always average over trajectories only \textit{after} first calculating the entanglement entropy or any other relevant quantities that can witness the phase transition.

\begin{figure}[tb]
 \centering
  \includegraphics[width=0.6\textwidth]{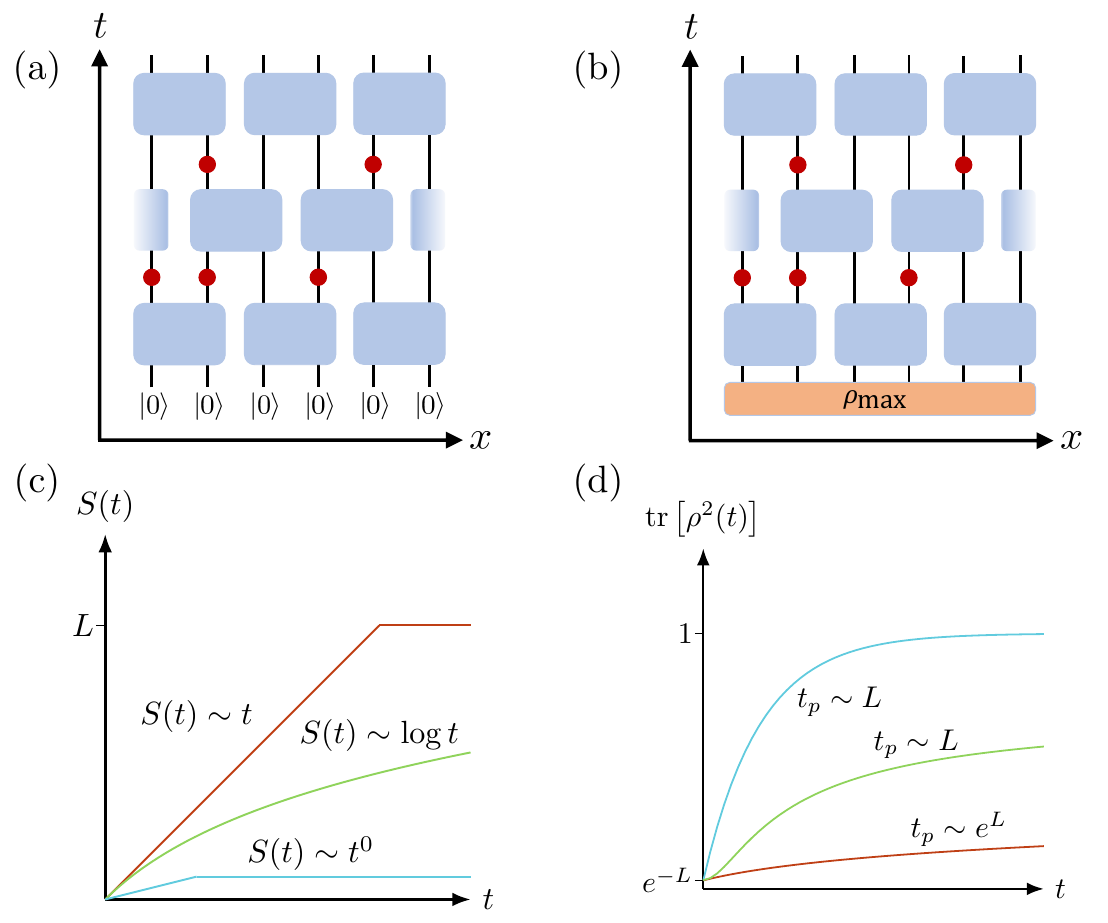}
 \caption{A monitored quantum circuit starting from a product state, consisting of rounds of nearest neighbor unitary gates (light blue rectangles), followed by rounds of projective measurements (red circles). Each spin independently has probability $p$ of being measured at a given time step. \textbf{(b):} The purification picture uses the same hybrid circuit as in \textbf{(a)}, but starts from the maximally mixed density matrix $\rho_{\mathrm{max}} = \mathds{1} / \tr{\mathds{1}}$. \textbf{(c):} Entanglement dynamics in monitored quantum circuits. For $p<p_{c}$ the entropy $S(t)$ grows ballistically before saturating to a volume-law. At criticality $p=p_{c}$ this is replaced by logarithmic growth, while in the area-law phase $p>p_{c}$ the entropy saturates in $\mathcal{O}(1)$ time. \textbf{(d):} Purification dynamics in monitored quantum circuits. Starting from the maximally mixed state, the purity $\tr{\rho^{2}}$ becomes $\mathcal{O}(1)$ in time $t_{p}\sim\mathcal{O}(L)$ in the purifying phase ($p>p_{c}$) and at criticality, whereas this takes time $t_{p}\sim\exp(\mathcal{O}(L))$ in the mixed phase $p<p_{c}$.}
  \label{Fig::Fig2}
\end{figure}

\subsection{Monitored quantum circuits}
As a minimal model, consider the quantum circuit shown in 
Fig.\ \ref{Fig::Fig2}~(a). 
This is a `hybrid' quantum circuit, consisting of rounds of nearest-neighbor unitary gates, followed by a round of measurements, where at each time step each spin has probability $p$ of being measured. For each spin that is measured, the probability of each outcome is determined by the Born rule, and the state is subsequently updated via a projection on to the corresponding state.

The case with the least structure, retaining only locality and unitarity, is when the unitary gates are drawn uniformly (according to the Haar measure) from the unitary group $\mathrm{U}(q^{2})$, where $q$ is the local Hilbert space dimension. Haar-random unitaries are highly chaotic, generating entanglement ballistically \cite{nahumQuantumEntanglementGrowth2017,vonkeyserlingkOperatorHydrodynamicsOTOCs2018} and forming unitary designs in polynomial time~\cite{brandaoLocalRandomQuantum2016}. For circuits with Haar-random gates on qubits, $q=2$, there is a phase transition in the steady state entanglement from volume-law to area-law at measurement probability $p_{c} \approx 0.17$~\cite{liQuantumZenoEffect2018,zabaloCriticalPropertiesMeasurementinduced2020}, where we can interpret $p$ as the density of measurements in spacetime, cf. Fig.\ \ref{Fig::Fig2}~(c). That this probability is nonzero may be surprising, given that naively local unitaries can only generate $\mathcal{O}(1)$ entanglement per time step, whereas local measurements can potentially destroy an extensive amount of entanglement. The resolution is that the local unitaries may `scramble' quantum information, such that the information about the state of a given subsystem is spread out over global degrees of freedom, with the effect that local measurements cannot learn much about the overall state of the system. This is very much in the spirit of quantum error correction, and indeed it has been argued that measurement-induced transitions can be viewed as phase transitions in the quantum channel capacity density $Q/N$, where the volume-law phase corresponds to $Q/N > 0$~\cite{gullansDynamicalPurificationPhase2020}. 

\subsection{Purification transition}
\label{sec:purification}

While the discussion in the previous section couched the measurement-induced transition in the language of an entanglement transition, it turns out that there is an alternative viewpoint in terms of a \textit{purification transition}~\cite{gullansDynamicalPurificationPhase2020}. Imagine keeping the same hybrid quantum circuit, but rather than using a product state as the initial state, using the maximally mixed state $\rho_{\mathrm{max}} = \mathds{1} / \tr{\mathds{1}}$ instead, as shown in 
Fig.\ \ref{Fig::Fig2}~(b). 
The action of the hybrid circuit will then be to purify the initial state, eventually reaching a pure state in the steady state. However, it turns out that there can be a phase transition in the \textit{time} $t_{p}$ taken for this purification --- from exponential to polynomial in system size, as shown in 
Fig.\ \ref{Fig::Fig2}~(d) --- and further that this transition seems to generically coincide with the entanglement transition for that class of hybrid quantum dynamics~\cite{gullansDynamicalPurificationPhase2020,luntMeasurementinducedCriticalityEntanglement2021}. Since in an experiment the effect of the environment is such that the state is generically mixed, unlike the pure state setting of the entanglement transition, the purification transition may provide a more robust lens through which to observe measurement-induced transitions in experiments~\cite{gullansDynamicalPurificationPhase2020}.

The purification picture also permits the introduction of an `order parameter' via coupling the system to an auxiliary system $R$~\cite{gullansScalableProbesMeasurementInduced2020}. In the simplest case, we can take $R$ to be a single qubit. We maximally entangle $R$ with a subset of the system at time $t_{0}$, and then let the system evolve in time --- the hybrid circuit acts only on the main system, so any dynamics in $R$ are induced solely through its entanglement with the system. The purification dynamics of $R$ then serve as a local probe of the measurement-induced phase structure~\cite{gullansScalableProbesMeasurementInduced2020}. By varying with which subsystem $R$ is entangled, and the time $t_{0}$ at which it is entangled, we can probe different critical exponents, as will be discussed further in \cref{sec:critical}.

\subsection{Transitions in the \renyi entropies}
\label{sec:renyi}

There are many quantities that can characterize the entanglement of a quantum state. Focusing on pure states, the most common choice is the von Neumann entropy, $S(A) = -\tr{\rho_{A}\log{\rho_{A}}}$, where $\rho_{A}$ is the reduced density matrix for subsystem $A$. But the von Neumann entropy turns out to be very difficult to measure in a large-scale experiment since it requires something like full state tomography~\cite{odonnellQuantumSpectrumTesting2015}, whose resource cost scales exponentially with system size. Happily, there is a related family of entropies, the \renyi entropies, which are more amenable to experimental access. Given a non-negative number $n$, the $n$-\renyi entropy $S_{n}(A)$ is defined as
\begin{equation}
	S_{n}(A) = \dfrac{1}{1-n} \log{\tr{\rho_{A}^{n}}},
	\label{eq:renyi_entropy}
\end{equation}
with the von Neumann entropy recovered in the limit $n \to 1$. For integer values of $n$ --- the easiest being $n=2$ --- the $n$-\renyi entropy can be measured in an `interferometry'-like experiment, where one prepares $n$ identical copies of the quantum state in question, and then measures the \renyi entropy through the expectation value of a certain observable. While this remains a considerable challenge, it has already been demonstrated experimentally~\cite{islamMeasuringEntanglementEntropy2015}, and so provides a possible route to accessing entanglement transitions in an experimental setting.

Given the experimental relevance of the \renyi entropies, it is natural to ask whether each \renyi entropy undergoes an entanglement transition at the same critical point. It turns out that while in general not all the \renyi entropies will transition at the same critical point, it follows from some basic properties of the \renyi entropies that many of them will in fact undergo the same entanglement transition. By differentiating with respect to $n$ the definition in \cref{eq:renyi_entropy} for the $n$-\renyi entropy, one can show that
\begin{equation}
    \dfrac{\diff S_{n}(A)}{\diff n} = \dfrac{-1}{(1-n)^{2}} D\left(\sigma \parallel \lambda\right),
\end{equation}
where $D( \sigma \parallel \lambda) = \sum_{i} \sigma_{i} \log(\sigma_{i}/\lambda_{i})$ is the relative entropy, taken between the probability distributions $\lambda = \{\lambda_{i}\}$, given by the eigenvalues of $\rho_{A}$, and $\sigma \equiv \{\lambda_{i}^{n} / \sum_{j} \lambda_{j}^{n}\}$. As a consequence of the non-negativity of the relative entropy, this implies that
\begin{equation}
    \dfrac{\diff S_{n}(A)}{\diff n} \leq 0.
\end{equation}
In other words, the $n$-\renyi entropies are \textit{non-increasing} as a function of $n$. In terms of the measurement-induced transition, this has the consequence that an area-law in the $n$-\renyi entropy implies an area-law in the $(m > n)$-\renyi entropies, and a volume-law in the $n$-\renyi entropy implies a volume-law in the $(m < n)$-\renyi entropies. Note however that we \textit{don't} necessarily have the converse, which would imply that $p_{c}$ is equal for all \renyi entropies.

However, we \textit{do} have the converse for $m, n > 1$. This is a consequence of the following inequality, valid only for $n>1$,
\begin{equation}
    S_{\infty} \leq S_{n} \leq \dfrac{n}{n-1} S_{\infty},
\end{equation}
which implies that all the $(n > 1)$-\renyi entropies must have the same scaling behaviors. This inequality can be proven by using monotonicity of the \renyi entropies and the fact that $S_{\infty}(A)$ is the largest eigenvalue of $\rho_{A}$.

\subsection{Analytically tractable limits}
\label{sec:analytic}

To determine the universality class of a phase transition, it is useful to have an analytic treatment. For measurement-induced transitions, such a treatment is currently available in a few select limits. A particularly simple treatment is available for the 0-\renyi entropy $S_{0}$, also known as the Hartley entropy. $S_{0}(A)$ simply gives the logarithm of the number of nonzero eigenvalues of the reduced density matrix $\rho_{A}$. It can be accessed through an adaptation of the `minimal cut' prescription for calculating the entanglement entropy~\cite{skinnerMeasurementInducedPhaseTransitions2019}, which first appeared in the context of random unitary circuits~\cite{nahumQuantumEntanglementGrowth2017}, and provides a `coarse-grained' picture for entanglement growth. The upshot is that the measurement-induced transition in the Hartley entropy in $d$-dimensional hybrid quantum circuits with Haar-random gates is in the universality class of $(d+1)$-dimensional percolation, where the extra dimension comes from the time direction of the quantum circuit. Furthermore, the critical measurement probability is precisely the bond percolation threshold for percolation on a lattice determined by the geometry of the quantum gates.

Percolation also appears in a different limit, corresponding to the transition in the $(n \geq 1)$-\renyi entropies strictly in the limit of infinite local Hilbert space dimension $q=\infty$. Here the connection to percolation is more subtle: it appears only in a replica limit $Q \to 1$ of a $Q!$-state Potts model obtained by averaging over unitary gates and measurements in a system where the gates are drawn from the Haar distribution over the unitary group~\cite{jianMeasurementinducedCriticalityRandom2020}. This treatment indicates that all the $(n \geq 1)$-\renyi entropies undergo a phase transition at the same critical point, consistent with the discussion in \cref{sec:renyi}. Interestingly, it turns out that the $q=\infty$ fixed point is unstable in the sense that finite $q$ explicitly breaks an emergent symmetry, allowing for the presence of an RG-relevant perturbation that drives the system away from the percolation fixed point. At the time of writing, the ultimate fate of that new fixed point is unknown, but remains a focus of current research.

\subsection{Critical properties of measurement-induced transitions}
\label{sec:critical}

In \cref{sec:analytic} we saw that in certain limits the measurement-induced transition was in the percolation universality class. However, both of these limits are somewhat unphysical: realistic quantum spin systems typically have finite local Hilbert space dimension, and the $0$-\renyi entropy can change discontinuously under arbitrarily small perturbations to the density matrix. Away from these limits, we turn to numerical probes to determine the relevant universality classes.

The critical exponents determining a universality class are typically obtained in numerics through finite-size scaling. In the rest of this section, we will mainly focus on numerics performed on \textit{Clifford circuits}, where the unitary gates are drawn uniformly from the Clifford group. Clifford circuits are convenient to look at in the context of measurement-induced entanglement transitions because they can be efficiently simulated classically~\cite{gottesmanHeisenbergRepresentationQuantum1998}, while still being able to rapidly generate entanglement. Their classical simulability is important in enabling sufficiently large system sizes that finite-size scaling provides good estimates of critical exponents. Furthermore, Clifford operations often turn out to be relatively easy to implement on current quantum computing devices, so hybrid Clifford circuits may be some of the first to be simulated in experiments. Box 1 outlines a method to classically simulate Clifford circuits using a graph-state based algorithm which is particularly suitable in the context of entanglement phase transitions.

\begin{figure}[tb]
 \centering
 \includegraphics[width = 0.85\textwidth]{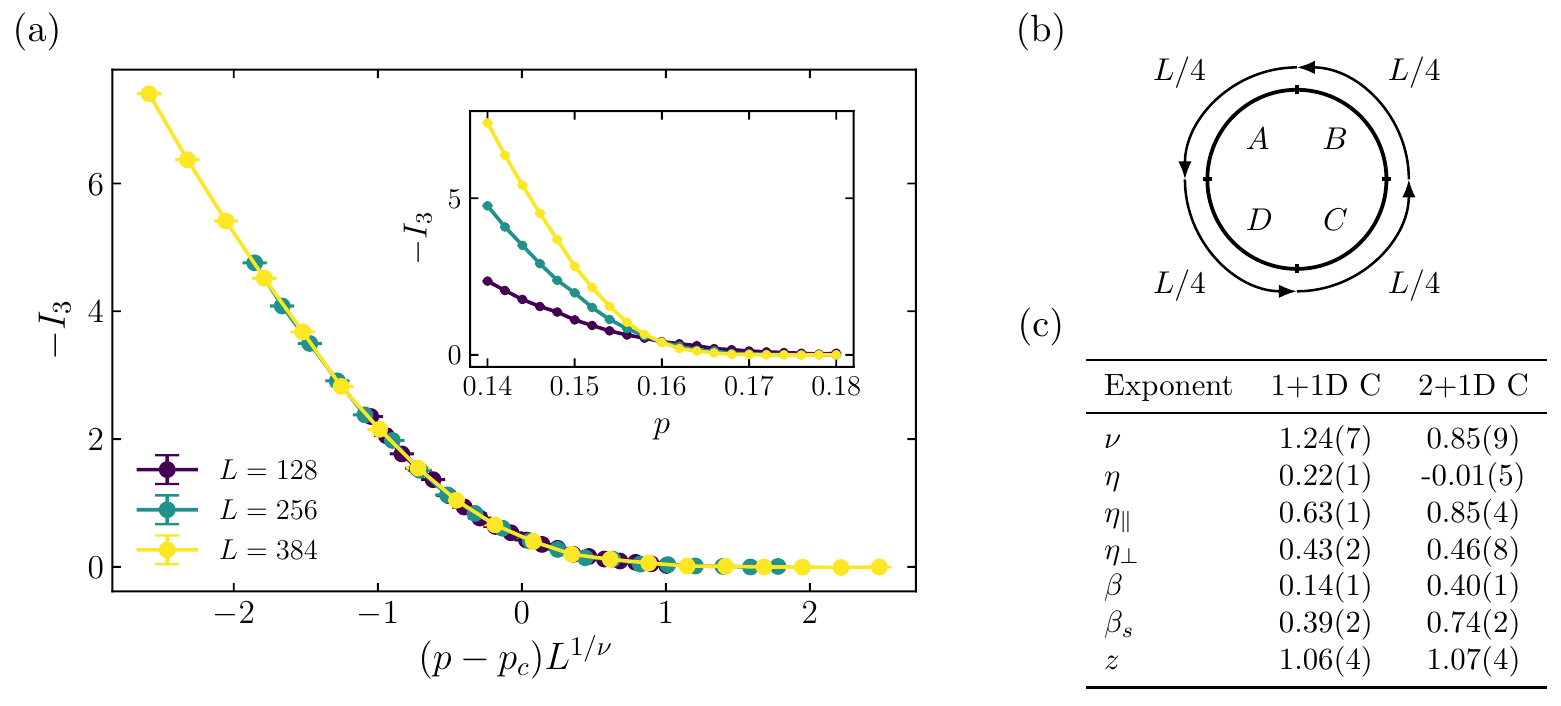}
 \caption{\textbf{(a):} The steady state tripartite information $I_{3}$ as a function of $(p-p_{c})L^{1/\nu}$, where $p_{c} \approx 0.158$ and $\nu \approx 1.33$. The inset shows the uncollapsed data. This dataset consists of $10^{4}$ circuit realizations. \textbf{(b):} The partition used for the tripartite information $I_{3}$ on a chain of length $L$ with periodic boundary conditions. \textbf{(c):} Critical exponents of the measurement-induced phase transition in 1+1D and 2+1D Clifford circuits, taken from Refs.~\cite{zabaloCriticalPropertiesMeasurementinduced2020,luntMeasurementinducedCriticalityEntanglement2021}.}
 \label{Fig::Fig3}
\end{figure}

To extract the critical point numerically, a natural quantity to look at in 1+1D is the half-chain entanglement entropy $S(L/2)$ (all \renyi entropies are equal in Clifford circuits). However, this has the disadvantage that it appears to scale logarithmically with system size $L$, as in a 1+1D conformal field theory~\cite{calabreseEntanglementEntropyQuantum2004}. This means that at the critical point the entanglement entropy for different system sizes does not coincide, so one cannot simply `read off' the critical point by looking at the data. Furthermore, if one attempts to extract the critical point $p_{c}$ and other critical exponents using standard finite-size scaling, these can have correlated errors depending on the extracted value of $p_{c}$. For a better estimator of the critical point, we turn to the \textit{tripartite information} $I_{3}$, defined as
\begin{equation}
	I_{3}(A : B : C) \equiv I_{2}(A : B) + I_{2}(A : C) - I_{2}(A : BC),
\end{equation}
where $I_{2}(A : B) \equiv S(A) + S(B) - S(AB)$ is the mutual information. One can show that for pure states, given a partition into four subsystems, the tripartite information of three of the subsystems does not depend on the choice of subsystems, so we will simply write $I_{3} \equiv I_{3}(A : B : C)$ from now on, with the partition in 
Fig.\ \ref{Fig::Fig3}~(b)
in mind. One can then argue using the `minimal cut' prescription of Ref.~\cite{skinnerMeasurementInducedPhaseTransitions2019} that for this choice of partition, $I_{3}$ cancels out the `boundary' contributions that give rise to the logarithmic scaling of the entanglement entropy at criticality~\cite{zabaloCriticalPropertiesMeasurementinduced2020}, resulting in the scaling
\begin{equation}
	I_{3}(p,L) \sim \begin{cases} \mathcal{O}(L), & p < p_{c}, \\ \mathcal{O}(1), & p = p_{c}, \\ 0, & p > p_{c}. \end{cases}
\end{equation}
Notably, this means that at the critical point $I_{3}(p_{c},L)$ should coincide for different system sizes $L$, providing a much more accurate estimator of the critical point. 
Figure \ref{Fig::Fig3}~(a) shows a data collapse of $I_{3}$ calculated in the steady state of 1+1D random Clifford circuits, which yields the estimate $p_{c} \approx 0.158$ for the critical point and $\nu \approx 1.33$ for the correlation length exponent. Notably, the latter is still very near the value of $\nu = 4/3$ for 2D percolation, despite the discussion in \cref{sec:analytic} that for finite local Hilbert space dimension (here $q=2$) we should expect the universality class to be distinct from percolation.

To measure other critical exponents, we turn to the `auxiliary system' method discussed in \cref{sec:purification}. For example, to extract the bulk exponent $\beta$, we initialize the system for a time $t_{0} = 2L$, and then entangle a reference qubit $R$ with the system qubit at position $x = L/2$, i.e.\ in the bulk. We then evolve the system for a further time $t = 2L$, and measure the entanglement entropy of $R$. Close to the phase transition, the circuit-averaged entropy should scale as $\overline{S(R)} \sim (p_{c} - p)^{\beta}$, reaching zero in the area-law phase. This yields the values $\beta = 0.14(1)$ in 1D~\cite{zabaloCriticalPropertiesMeasurementinduced2020} and $\beta = 0.40(1)$ in 2D~\cite{luntMeasurementinducedCriticalityEntanglement2021}. Again both of these values are reasonably close to the corresponding values for 2D and 3D percolation, $\beta_{2D} = 5/36 \approx 0.139$ and $\beta_{3D} \approx 0.43$. However, there is increasing evidence that, despite these similarities, the measurement-induced transition in Clifford circuits is indeed in a different universality class to percolation~\cite{zabaloCriticalPropertiesMeasurementinduced2020,luntMeasurementinducedCriticalityEntanglement2021,liConformalInvarianceQuantum2021,liStatisticalMechanicsModel2021}.

We can also extend this method to measure correlation function exponents. We introduce two auxiliary systems, $R_{1}$ and $R_{2}$, which are entangled with separate subsystems at a time $t_{0}$, and then subject to dynamics again only through their entanglement with the main system. We can extract correlation function exponents by studying their mutual information $I_{2}(R_{1} : R_{2})$, which serves as an upper bound on all connected correlation functions of local observables~\cite{wolfAreaLawsQuantum2008}. For example, the exponent $\eta$ governs the power-law decay of bulk correlation functions
\begin{equation}
	\left|\langle A_{r} B_{0} \rangle - \langle A_{r} \rangle \langle B_{0} \rangle\right| \sim \dfrac{1}{r^{d-2+\eta}},
	\label{eq:critical_correlation_function}
\end{equation}
for local observables $A_{r}$ and $B_{0}$ supported at sites $r$ and $0$ respectively. We can extract this exponent by initializing the system for time $t_{0} = 2L$, entangling auxiliary qubits $R_{1}$ and $R_{2}$ with the bulk qubits at positions $x = L/4$ and $x = 3L/4$, and then letting the system evolve for a further time $t=2L$. The mutual information should then decay as $I_{2}(R_{1} : R_{2}) \sim 1/L^{\eta}$ in 1+1D, so we can extract $\eta$ by optimizing for a data collapse of $L^{\eta}I_{2}(R_{1} : R_{2})$ for different system sizes $L$. We could also extract the corresponding surface correlation length exponents $\eta_{\bot}$ and $\eta_{\parallel}$ by choosing to entangle one or both of $R_{1}$ and $R_{2}$ to a surface qubit instead of a bulk qubit. In 
Fig.\ \ref{Fig::Fig3}~(c) we reproduce from Refs.~\cite{zabaloCriticalPropertiesMeasurementinduced2020,luntMeasurementinducedCriticalityEntanglement2021} the current numerical estimates of the critical exponents of the measurement-induced transition in 1+1D and 2+1D Clifford circuits, which were determined using the methods described here.

\clearpage

\colorlet{shadecolor}{blue!80}
\begin{shaded}
\noindent 
 \textcolor{white}{\bf Box 1 $|$ SIMULATING CLIFFORD CIRCUITS WITH GRAPH STATES}
\end{shaded}
\vspace{-2mm}
\colorlet{shadecolor}{blue!10}
\begin{shaded}
\noindent

The Clifford group $\mathcal{C}_{n}$ on $n$ qubits is defined as the group that preserves the corresponding Pauli group $\mathcal{P}_{n}$ under conjugation, modulo global phases. To be concrete, we focus on models with nearest-neighbor unitary gates drawn uniformly from the two-qubit Clifford group $\mathcal{C}_{2}$, akin to the sketch in 
Fig.\ \ref{Fig::Fig2}~(a), and with projective measurements in the $\sigma^{z}$ basis. The two-qubit Clifford group can be generated by the gates $\{H, S, CZ\}$, where $CZ$ is a two-qubit controlled-$Z$ gate, and $H$ and $S$ are the single-qubit Hadamard and phase gates given in the computational basis by
\begin{equation}
	H = \dfrac{1}{\sqrt{2}} \begin{pmatrix} 1 & 1 \\ 1 & -1 \end{pmatrix}, \qquad S = \begin{pmatrix} 1 & 0 \\ 0 & i \end{pmatrix}.
	\label{eq:clifford_gates}
\end{equation}
Ignoring overall phases, the resulting group has 11520 elements, which is sufficiently small that it can be hardcoded as a lookup table decomposing each group element into a product of the generators. This allows for simulations of random Clifford circuits using a graph-state based algorithm~\cite{andersFastSimulationStabilizer2006}, which makes use of the fact that all stabilizer states resulting from a Clifford circuit can be written as a graph state, up to the action of some one-qubit Clifford gates~\cite{heinMultipartyEntanglementGraph2004}. Alternative simulation methods exist, most notably one based on storing a `stabilizer tableau' of the stabilizers fixing the quantum state, and then updating the tableau based on the action of Clifford gates~\cite{aaronsonImprovedSimulationStabilizer2004}. However, the graph-state method is particularly well suited to the study of measurement-induced transitions, because its time complexity scales with the typical vertex degree of the graph storing the quantum state. This degree is generically reduced by the presence of measurements, such that it is possible to simulate very large system sizes in the area-law phase and near the critical point.

\textbf{Graph states and stabilizer states}

The graph-state algorithm relies on a fortunate connection between graph states and stabilizer states, the latter being the states generated by Clifford circuits acting on the $\ket{0}^{\otimes N}$ initial state. Namely, all stabilizer states can be written as a graph state, up to the action of some one-qubit Clifford gates~\cite{heinMultipartyEntanglementGraph2004}. Given a graph $\mathcal{G} = (V,E)$, the graph state $\ket{\mathcal{G}}$ is defined by
\begin{equation}
	\ket{\mathcal{G}} = \left(\prod_{(i,j) \in E} CZ_{ij} \right) \ket{+}^{\otimes N}, 
\end{equation}
where $\ket{+} = (\ket{0}+\ket{1})/\sqrt{2}$, and $N$ is the total number of qubits. That is, starting from the initial state $\ket{+}^{\otimes N}$, we perform a controlled-$Z$ gate between all pairs of qubits $(i,j)$ connected by an edge in the graph. Thus, to represent the graph state $\ket{\mathcal{G}}$ we just need to store the graph $\mathcal{G}$, which only takes $\mathcal{O}(N^{2})$ memory. We can then write any stabilizer state as
\begin{equation}
	\ket{\mathcal{G} ; \{C_{i}\}} = \left(\bigotimes_{i=1}^{N} C_{i} \right) \ket{\mathcal{G}},
\end{equation}
where the $C_{i}$ are drawn from the one-qubit Clifford group $\mathcal{C}_{1}$. This group has only 24 elements up to phase, so this additional information is only an $\mathcal{O}(N)$ overhead.

The entanglement in a graph state is completely fixed by the graph $\mathcal{G}$. Suppose we wanted to calculate the entanglement entropy of a subsystem $A$. Partitioning the adjacency matrix $\Gamma$ of $\mathcal{G}$ into the block form
\begin{equation}
	\Gamma = \begin{pmatrix}
				\Gamma_{AA} & \Gamma_{AB} \\
				\Gamma_{AB}^{\mathrm{T}} & \Gamma_{BB}
	\end{pmatrix},
\end{equation}
where $B$ is the complement of $A$, the entanglement entropy $S_{A}$ is given by
\begin{equation}
	S_{A} = \mathrm{rank}_{\mathbb{F}_{2}}(\Gamma_{AB}),
\end{equation}
the rank over the binary field $\mathbb{F}_{2}$ of the subadjacency matrix $\Gamma_{AB}$ characterizing connections between $A$ and $B$~\cite{heinEntanglementGraphStates2006}.

\clearpage

\textbf{Implementing Clifford operations}

\begin{wraptable}{r}{0.4\textwidth}
	\begin{tabular}{cc}
		\toprule
		Clifford operation & Time complexity\\
		\midrule
		One-qubit gate & $\Theta(1)$ \\
		Two-qubit gate & $\mathcal{O}(d^{2})$ \\
		Pauli $Z$ measurement & $\mathcal{O}(d)$\\
		\bottomrule
	\end{tabular}
	\caption{Time complexity of different Clifford operations. $d$ is the maximum vertex degree of the qubits involved in the gate, whose scaling with system size depends on the entanglement phase.}
	\label{tab:clifford_operations}
\end{wraptable}

Note that there is some `gauge freedom' in writing stabilizer states this way, so different combinations $(\mathcal{G}; \{C_{i}\})$ can correspond to the same quantum state. This freedom turns out to be useful in implementing two-qubit Clifford operations on the quantum state. Here we will summarize how one performs Clifford gates and Pauli measurements on stabilizer states represented in this way --- for the full details we refer the reader to Ref.~\cite{andersFastSimulationStabilizer2006}. The relevant time complexities of the different operations are shown in \cref{tab:clifford_operations}.

One-qubit Clifford gates are trivial to perform since they leave the graph invariant --- we merely have to update the one-qubit Clifford $C_{i}$ corresponding to the site $i$ of the gate, which takes $\Theta(1)$ time. This can be done with a lookup table of size $|\mathcal{C}_{1}|^{2} = 24^{2}$. Two-qubit Clifford gates are more technical, since they involve both the graph and the one-qubit Cliffords. We only need to focus on implementing $CZ$ gates, since this is the only two-qubit gate in the generating set of the two-qubit Clifford group. There are two cases in implementing $CZ$ on a given pair of qubits $(i,j)$. The easy case is if the corresponding one-qubit Cliffords $C_{i}$ and $C_{j}$ commute with $CZ_{i,j}$. In this case, the $CZ$ leaves the one-qubit Cliffords unchanged, and toggles the edge $(i,j)$ in the graph. The harder case is if $CZ_{i,j}$ does not commute with the one-qubit Cliffords, and it is here that the `gauge freedom' becomes useful. The goal is to move to a gauge where the problem is reduced to the easy case previously described. We do this using an operation called `local complementation', which toggles all the edges in the subgraph induced by the neighborhood of a given vertex, say $i$, and then modifies the one-qubit Cliffords of site $i$ and its neighbors. In this way, one can obtain $C_{i}$ and $C_{j}$ which commute with $CZ_{i,j}$. The local complementation is the dominant cost, taking time $\mathcal{O}(d^{2})$, where $d$ is the maximum vertex degree of qubits $i$ and $j$. This scaling with $d$ rather than system size $N$ means the runtime of this algorithm depends strongly on the connectedness of the graph, and hence, roughly speaking, that less entangled states are quicker to simulate. This is especially pertinent in the context of entanglement phase transitions. In principle, $d$ can be as large as $\mathcal{O}(N)$, and this scaling is relevant in the volume-law phase. However, in the area-law phase and near the critical point, the typical vertex degree can be $\mathcal{O}(1)$, so that even two-qubit gates are easy to implement.

Finally we describe how to perform single-site Pauli measurements. The basic idea is to reduce the measurement on the stabilizer state $\ket{\mathcal{G} ; \{C_{i}\}}$ to a measurement on the underlying graph state $\ket{\mathcal{G}}$ without the one-qubit Cliffords. Suppose we measure Pauli $P_{a}$ on site $a$, with outcome $\lambda \in \{ \pm 1 \}$. The stabilizer state will be updated to
\begin{align}
	\dfrac{\mathds{1}+\lambda P_{a}}{2} \ket{\mathcal{G} ; \{C_{i}\}} &= \left( \prod_{b \in V \setminus \{a\}} C_{b} \right)  \dfrac{\mathds{1}+\lambda P_{a}}{2} C_{a} \ket{\mathcal{G}} \\
	&= \left( \prod_{b \in V \setminus \{a\}} C_{b} \right) C_{a}  \dfrac{\mathds{1}+\lambda C_{a}^{\dagger} P_{a} C_{a}}{2} \ket{\mathcal{G}},
\end{align}
where we inserted a factor of $\mathds{1} = C_{a} C_{a}^{\dagger}$ in the last step. Since $C_{a}$ is a Clifford operator and preserves spectra, $P_{a}^{'} \equiv C_{a}^{\dagger} P_{a} C_{a} \in \{\pm X_{a}, \pm Y_{a}, \pm Z_{a}\}$. Hence the effect of measuring Pauli $P_{a}$ on the stabilizer state $\ket{\mathcal{G} ; \{C_{i}\}}$ can be modelled by measuring Pauli $P_{a}^{'}$ on the graph state $\ket{\mathcal{G}}$. Relegating full details of graph state measurements to Refs.~\cite{heinMultipartyEntanglementGraph2004,andersFastSimulationStabilizer2006}, we note that Pauli $Z$ measurements are particularly simple: they simply remove all edges from the graph connected to the measured site, which takes $\mathcal{O}(d)$ time.

\end{shaded}

\subsection{Entanglement transitions in experiments}

\subsubsection{Scalability issues}

As discussed in \cref{sec:renyi}, it is very difficult to experimentally measure the von Neumann entropy, but this can possibly be ameliorated by instead measuring a \renyi entropy, such as the 2-\renyi entropy $S_{2}$ (related to the purity via $\tr{\rho^{2}} = \exp[-S_{2}]$). However, before one can get to that point one has to reliably and repeatedly prepare a given steady state for which the entropy is to be calculated, and it is here that the probabilistic nature of the measurements can present an issue of scalability. Even if we fix the measurement locations for simplicity, their random outcomes mean that for $T$ rounds of measurements with probability $p$, in a system of $N$ spins with local Hilbert space dimension $q$, the number of repetitions required to get $\mathcal{O}(1)$ samples of a given trajectory is of the order $q^{p N T}$, which is exponential in the system size. To make things worse, the equilibration time $T$ is typically at least linear in the system size $N$, so the sampling overhead can be \textit{doubly} exponential in $N$. In principle, this presents a severe barrier to scaling up these experiments to the large system sizes where phase transitions are most apparent.

There are several options to try to avoid this particular issue of postselection. One option is to relate entanglement entropy to a quantity which can be more easily measured. In particular, this is possible by relating a given hybrid quantum circuit to its \textit{spacetime dual}. Given a unitary matrix $U$ acting on a pair of spins, we denote its matrix elements by $[U]_{i_{\mathrm{in}} j_{\mathrm{in}}}^{i_{\mathrm{out}} j_{\mathrm{out}}}$. This matrix is unitary in the `time' direction, expressed by the condition $\sum_{kl} [U]_{i_{\mathrm{in}} j_{\mathrm{in}}}^{kl} [U^{*}]_{i_{\mathrm{out}} j_{\mathrm{out}}}^{kl} = \delta_{i_{\mathrm{in}} j_{\mathrm{in}}, i_{\mathrm{out}} j_{\mathrm{out}}}$. However, we could decide to swap the space and time directions: viewing the unitary $U$ as a tensor with four legs, two for input and two for output, we form an associated tensor $\widetilde{U}$, where one of the original output legs of $U$ becomes an input leg of $\widetilde{U}$, and one of the original input legs of $U$ becomes an output leg of $\widetilde{U}$. The resulting $\widetilde{U}$ is generically nonunitary, and can be interpreted as a unitary gate followed by a weak measurement~\cite{nappEfficientClassicalSimulation2019}. Thus by taking the spacetime dual we have a relation between a hybrid quantum circuit and an associated unitary circuit. It turns out that one can relate the 2-\renyi entropy in the hybrid circuit to a correlation function in the unitary circuit~\cite{ippolitiPostselectionFreeEntanglementDynamics2021}, which is typically much easier to measure in an experiment.

Another option is to consider a modification of the measurement protocol, where instead of having random measurements with outcomes distributed according to the Born rule, we consider a non-unitary but deterministic time-evolution, such as that resulting from a non-Hermitian Hamiltonian~\cite{gopalakrishnanEntanglementPurificationTransitions2021}. In effect one can think of this as being similar to `forcing' certain measurement outcomes.~\cite{nahumMeasurementEntanglementPhase2021}. Non-Hermitian Hamiltonians can emerge quite naturally in certain open quantum systems undergoing continuous measurement. However, it is worth noting that the measurement-induced transition in these systems may be somewhat different in character to that in random quantum circuits with Born rule projective measurements---in replica treatments of the phase transition (see \cref{sec:analytic}), the Born rule factor necessitates an additional replica compared to the case of forced measurements~\cite{jianMeasurementinducedCriticalityRandom2020}, so the latter in fact appears to be more like closely related transitions in random tensor networks~\cite{vasseurEntanglementTransitionsHolographic2019}. Additionally, the lack of quenched randomness with forced measurements may give rise to qualitatively different behavior of the entanglement domain walls that arise in statistical mechanical models describing these transitions.

\subsubsection{Measurement-induced transition in a trapped-ion experiment}

Ref.~\cite{noel2021observation} presented the first experimental observation of a measurement-induced transition. They studied a chain of 13 trapped \ce{^{171}Yb+} ions, with relatively high gate fidelities of 99.96\% for single-qubit gates and 98.5-99.3\% for two-qubit gates. Instead of directly measuring the entanglement entropy, they focused on the measurement-induced transition in the purification picture, as discussed in \cref{sec:purification}, since this is typically more robust to experimental imperfections, and also made use of the following simplification. In this system certain operations are easier to perform than others: the `native' two-qubit gate they employ is an `Ising' XX gate of the form $U(\theta) = \exp(-i \theta \sigma_{i}^{x}\sigma_{j}^{x})$, where the value of $\theta$ is controlled by the duration of a control pulse. The use of this gate implies that if measurements are performed in the $\sigma^{x}$ basis then we will quickly approach a $\sigma^{x}$-basis product state, since the XX Ising gates cannot generate entanglement from $\sigma^{x}$ basis states. Exploiting this fact, the authors consider a modification of the usual purification protocol, fixing the overall measurement probability $p$, but performing the measurements in two different bases, the $\sigma^{x}$ and $\sigma^{z}$ bases. There is an additional parameter $p_{x}$ that controls the probability that a given measurement is performed in the $\sigma^{x}$ basis. With $p$ fixed to a small value to limit the overall number of measurements, there is a phase transition as a function of $p_{x}$ between the slow- and fast-purifying phases.

\section{Random circuits on noisy-intermediate scale quantum 
devices}\label{Sec::NISQ}

As outlined so far, random circuits with different designs 
have proven extremely useful to understand the properties of quantum 
many-body systems and the dynamics of entanglement. Recently, random 
circuits have found a new application in quantum information science. In 
particular, they have been used to demonstrate ``quantum 
supremacy'' or, in other words, a quantum computational advantage, which refers 
to the fact that a quantum device can perform a task that is unfeasible for its 
classical counterpart \cite{Preskill2018}. Such a demonstration of a 
quantum advantage is especially nontrivial given the eponymous 
noise of today's NISQ devices. This is not least the reason why, as a 
computational problem to achieve ``quantum supremacy'', it was chosen to 
sample from the output probability distribution of a random circuit, which is 
believed to 
be a hard task for classical supercomputers. In Sec.\ 
\ref{Sec::Circuit::Sampling}, we will 
provide an introduction to the main ideas of random-circuit sampling. 
Moreover, in Sec.\ \ref{Sec::Hydro_on_QC}, we will discuss that 
random circuits on NISQ devices are not just abstract tools to outperform 
classical computers, but are useful in a wider range of applications. To 
this end, we will leverage the concept of quantum typicality (see Box 2) and 
focus on the 
results 
reported in Ref.\ \cite{Richter_2021}.      

\subsection{Random-circuit sampling for achieving a quantum computational 
advantage}\label{Sec::Circuit::Sampling}

Random circuits are natural candidates when striving to outperform classical 
computers. On one hand, random circuits seem more appealing than some 
arithmetic calculation in view of the noise of today's NISQ devices. On the 
other hand, 
these circuits are challenging for classical machinery due to the lack of 
structure and the quick generation of entanglement.  
The main idea of random-circuit sampling is as follows \cite{Boixo_2018, 
Arute_2019}. Given a random circuit 
${\cal R}$, apply ${\cal R}$ to the $L$ qubits of the system and  
measure all qubits afterwards, which yields a bitstring
$\ket{k} = \ket{010110\cdots}$ according to Born's rule with  
probability $z = |\langle k|{\cal R}|0\rangle|^2$. Repeating this experiment 
many times then yields a set of bitstrings $\ket{k_1}, \ket{k_2}, \ket{k_3}, 
\dots$, which allows to draw conclusions on the underlying probability 
distribution. In particular, for random circuits that are sufficiently deep, it 
is expected that the resulting state is Haar-random \cite{Boixo_2018, 
Oliveira2007, Weinstein2008, Emerson2003, Harrow2009, Znidaric2007}, i.e., it is 
a realization 
drawn from the uniform distribution on the $2D$-dimensional unit sphere, where 
$D = 2^L$ is the dimension of the Hilbert space. In practice, 
this implies that the complex coefficients $c_k$ of the 
wave function ${\cal R}\ket{0} = \sum_k c_k \ket{k}$ are drawn from a Gaussian 
distribution and that the probability 
distribution of $z = |c_k|^2$ follows as \cite{Boixo_2018, Jin_2021}, 
\begin{equation}\label{Eq::PT}
 p(z) = (D-1)(1-z)^{D-2} \approx e^{-Dz}\ , 
\end{equation}
where the approximation on the right hand side of Eq.\ \eqref{Eq::PT} applies 
to large Hilbert-space dimensions $D$, and is sometimes referred to as 
Porter-Thomas law \cite{Porter1956}. 

In order to demonstrate a quantum computational advantage using random-circuit 
sampling, several 
key points have to be addressed \cite{Jin_2021}. First of all, it is 
important to 
execute the random circuit on a large system that surpasses the capacities of 
classical computers (currently the best supercomputers can evaluate 
circuits with $L \approx 45$ \cite{DeRaedt2019}). However, for such large 
systems, the 
corresponding Hilbert space is huge and reconstructing the probability 
distribution $p(z)$ would require an impracticably high number of bitstrings 
(experimental repetitions). Moreover, how can one verify that the quantum 
device indeed samples from the correct probability distribution if classical 
verification is impossible? An important step to overcome these difficulties 
has been achieved by the introduction of cross-entropy benchmarking (XEB) 
\cite{Boixo_2018}. To 
understand XEB, consider an ideal (quantum) computer that 
can execute ${\cal R}$ and thus provides the probabilities $p(\ket{k})$ of all 
$2^L$ bitstrings. Now, let a NISQ device execute ${\cal R}$ and we collect 
$N$ samples ${\cal S} = \lbrace \ket{k_1}, \dots, \ket{k_N} \rbrace$. If $N$ is 
large enough, we could reconstruct the corresponding probability distribution 
$p_\text{NISQ}(\ket{k})$ and compare it with the ideal distribution 
$p(\ket{k})$ using standard statistical tools. However, even for a smaller 
number of samples, it was shown that conclusions on the quality of the NISQ 
device can drawn based on an approximation of the cross entropy 
\cite{Boixo_2018, Jin_2021},   
\begin{equation}\label{Eq::XEB}
 S_\text{XEB} = -\sum_{k=1}^D p_\text{NISQ}(\ket{k})\log 
p(\ket{k}) \approx   -\frac{1}{N}\sum_{\ket{k} \in {\cal S}} \log 
p(\ket{k})\ , 
\end{equation}
where the second part results from replacing the sum over all $D$ 
bitstrings by the sample average justified by the central limit theorem. 
In practice, the probabilities $p(\ket{k})$ to perform XEB are obtained  
numerically by simulating the quantum circuit on a supercomputer. 
Obviously, this is not possible in the regime of system sizes and circuit 
depths where a quantum computational 
advantage occurs. Nevertheless, various sophisticated techniques have been 
developed (e.g., considering elided circuits with fewer entangling gates 
\cite{Arute_2019}) to 
estimate the cross entropy in Eq.\ \eqref{Eq::XEB} also 
in such cases. For more details on XEB, we refer to \cite{Boixo_2018, 
Arute_2019, Jin_2021}. 

A quantum computational advantage based on random-circuit 
sampling has been announced for the first time using Google's 
NISQ device {\it Sycamore} \cite{Arute_2019}. In the experiment, the Josephson 
junction-based processor executed a random circuit consisting of layers of 
one-qubit and two-qubit gates. We will discuss such types of circuits in more 
detail in Sec.\ \ref{Sec::Hydro_on_QC}. Due to imperfections in the execution, 
it was found in \cite{Arute_2019} that the cross entropy decays exponentially 
with the 
number of qubits and the depth of the circuit. Nevertheless, a nonzero 
$S_\text{XEB}$ was established even for the largest circuits with $53$ qubits 
and $20$ cycles of one-qubit and two-qubit gate layers, implying that the 
processor samples from a nontrivial probability distribution. 
In particular, according to Google's estimate at that time, a classical 
supercomputer would require 10,000 years for the same task (albeit significantly 
lower estimates have been suggested subsequently \cite{Gray2021, Huang2020}). In 
more recent experiments, 
a team from China used their superconducting quantum processor {\it 
Zuchongzhi} to perform random-circuit sampling for even larger qubit numbers 
and circuit depths, which impressively substantiate the fact that a quantum 
computational advantage has indeed been achieved \cite{Wu2021, Zhu2021}. 
Eventually, it is important 
to note that a quantum computational 
advantage is of course a moving target since both quantum and classical 
hardware as well as algorithms are expected to further improve in the future 
\cite{Zlokapa2020}.

\subsection{Applications of random circuits in quantum 
many-body 
physics}\label{Sec::Hydro_on_QC}

While sampling the outcome of random circuits is arguably a rather abstract 
task, it is an  
intriguing question to what extent the current capabilities of NISQ devices can 
be 
leveraged to study a wider class of problems. In this context, we 
specifically refer to questions in quantum many-body dynamics, which are 
generally challenging for classical computers due to the exponentially large 
Hilbert space. Simulations on NISQ devices might provide an 
opportunity to tackle these challenges and, moreover, to open up novel 
directions of research. In particular, it is important to note that today's 
NISQ 
devices, with of the order of $50$ qubits, already operate in Hilbert spaces 
that are competitive to or even go beyond what is possible with the best 
supercomputers, as also demonstrated recently \cite{Mi2021, Mi2021_2, 
Satzinger2021} \\
 
\begin{figure}[tb]
 \centering
 \includegraphics[width = 0.9\textwidth]{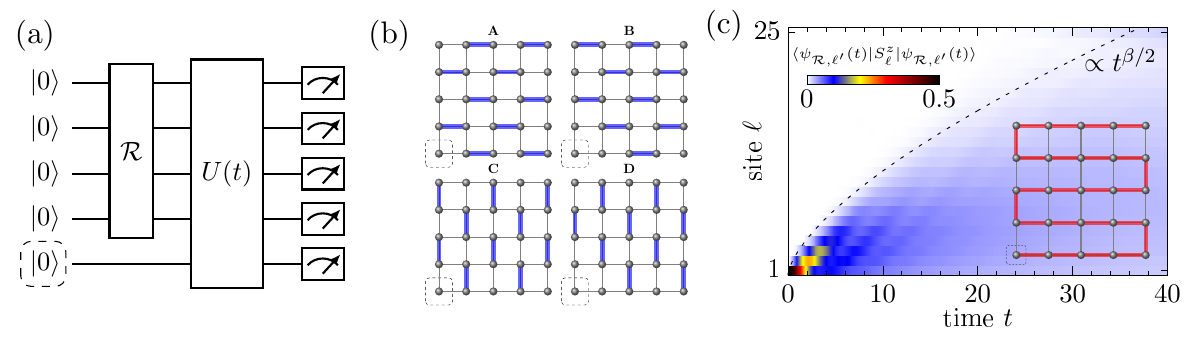}
 \caption{{\bf (a)} Algorithm to simulate dynamical spin-spin correlation 
function. A random circuit ${\cal R}$ acts on $L-1$ qubits (except 
site $\ell'$), followed by a time evolution $U(t)$ on all $L$ sites. 
Measurement on site $\ell$ then yields $C_{\ell,\ell'}(t)$. {\bf (b)} We here 
consider a two-dimensional geometry. Similar to \cite{Boixo_2018, Arute_2019}, 
${\cal R}$ consists of 
layers of one-qubit and two-qubit gates, cf.\ Fig.\ 
\ref{Fig::Random}~(c). The patterns {\bf A}-{\bf D} indicate the position of the 
two-qubit gates in different cycles. {\bf (c)} For reference site $\ell' = 1$, 
the 
expectation value $\bra{\psi_{{\cal R},\ell'}(t)}S_\ell^z\ket{\psi_{{\cal 
R},\ell'}(t)}$ yields $2C_{\ell,\ell'}(t)$. Data is shown for the Heisenberg 
chain with $L = 25$, where the dynamics of the one-dimensional 
system can be simulated by considering a snake-like path through the 
two-dimensional grid. Figure adapted from Ref.\ 
\cite{Richter_2021}.}
 \label{Fig::Scheme}
\end{figure}

In this section, we discuss a recent proposal to simulate hydrodynamics in 
quantum many-body systems on NISQ devices using random circuits 
\cite{Richter_2021}, which emphasizes that random circuits 
are not just useful to demonstrate ``quantum supremacy'', but they in fact form 
tailor-made building blocks to study questions in quantum many-body physics. The 
two key points in this context are that (i) random circuits swiftly generate 
random highly entangled quantum states, and (ii) the properties of such random 
 states can be exploited for efficient simulations by relying on the concept 
of quantum typicality \cite{Lloyd1988, Goldstein2006, Popescu2006, Reimann2007} 
(see Box 2).   
More specifically, Ref.\ \cite{Richter_2021} presented a scheme to 
compute the infinite-temperature spatiotemporal correlation function 
$C_{\ell,\ell'}(t)$, 
\begin{equation}\label{Eq::Korrel}
 C_{\ell,\ell'}(t) = \frac{\text{tr}[S_\ell^z(t) S_{\ell'}^z]}{2^{L}}\ , 
\end{equation}
where $S_\ell^z$ is a spin-$1/2$ operator at lattice site $\ell$, $S_\ell^z(t) 
= e^{i{\cal H}t}S_\ell^z e^{-i{\cal H}t}$ is the time-evolved operator with 
respect to a Hamiltonian ${\cal H}$ (although, in principle, evolutions with 
respect to other unitaries $U(t)$ are conceivable as well), and 
$L$ denotes the numbers of spins (qubits). 
The algorithm to simulate $C_{\ell,\ell'}(t)$ is sketched in Fig.\ 
\ref{Fig::Scheme}~(a). First, all qubits are initialized in the $\ket{0}$ 
state. (We identify $\ket{0} \equiv \ket{\uparrow}$ and $\ket{1} \equiv 
\ket{\downarrow}$ in the following.) A random circuit ${\cal R}$ then acts on 
$L-1$ qubits (except for site $\ell'$). The resulting state $\ket{\psi_{{\cal 
R},\ell'}} = \ket{0}_{\ell'} \otimes {\cal R} \ket{0}^{\otimes L-1}$ is evolved 
in time, $\ket{\psi_{{\cal 
R},\ell'}(t)} = e^{-i{\cal H}t}  \ket{\psi_{{\cal 
R},\ell'}}$, and the measurement at site $\ell$ yields 
the spatiotemporal correlation function according to (see \cite{Richter_2021} 
for a derivation), 
\begin{equation}\label{Eq::Korrel_2}
 C_{\ell,\ell'}(t) = \frac{1}{2}\bra{\psi_{{\cal 
R},\ell'}(t)}S_\ell^z\ket{\psi_{{\cal 
R},\ell'}(t)} + {\cal O}(2^{-L/2})\ 
, 
\end{equation}
where the statistical error of the quantum-typicality approximation vanishes 
exponentially with increasing system size \cite{Jin_2021}. In this context, 
quantum typicality 
can be interpreted as a form of ``quantum parallelism'' 
\cite{Schliemann_2002, _lvarez_2008}, as the time evolution of a single random 
state $\ket{\psi_{{\cal 
R},\ell'}}$ already captures the full ensemble average. This typicality-based 
approach to simulate dynamical two-point correlation functions such as 
$C_{\ell,\ell'}(t)$ on a quantum computer is complementary to other established 
schemes \cite{Terhal2000, Somma2002, Pedernales2014}, and operates without an 
overhead of bath or ancilla qubits. Crucially, 
it is a direct extension of the random-circuit experiments already realized on 
today's NISQ devices \cite{Arute_2019, Wu2021, Zhu2021}.

\begin{figure}[tb]
 \centering
 \includegraphics[width = 0.8\textwidth]{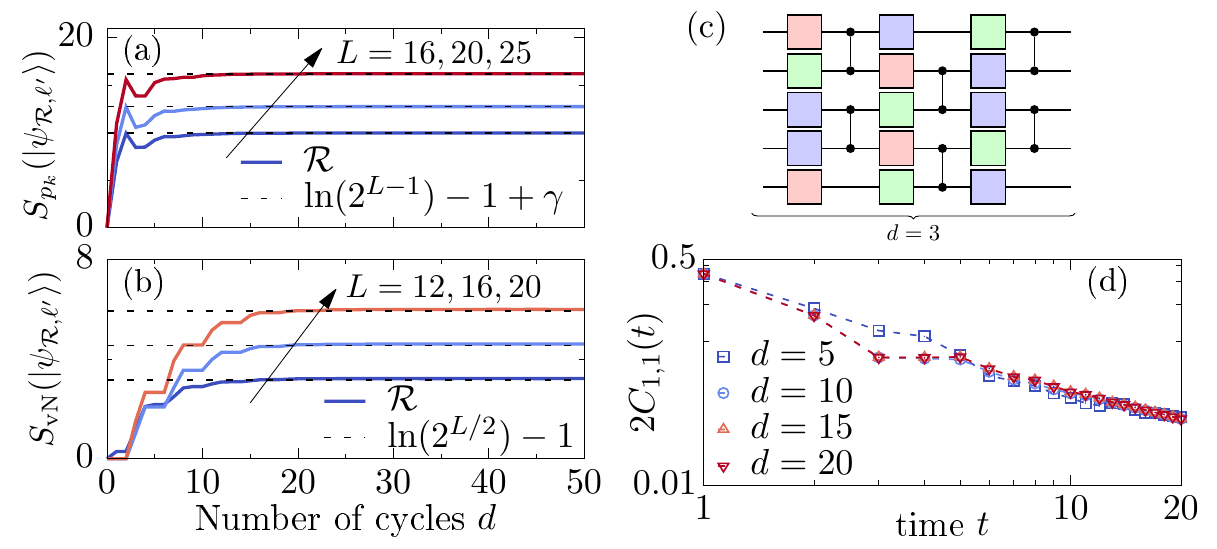}
 \caption{[{\bf (a),(b)}] Buildup of randomness of $\ket{\psi_{{\cal 
R},\ell'}}$ 
versus the depth $d$ of ${\cal R}$ [cf.\ panel (c)], 
measured by (a) $S_{p_k}$ and (b) $S_\text{vN}$. Both quantities reach their 
corresponding random-state values already at moderate $d$. The 
displayed values of $L$ correspond to two-dimensional geometries $4\times 3$, 
$4\times 4$, $5\times 4$, and $5\times 5$. Data is averaged over $100$ 
realizations of ${\cal R}$. {\bf (c)} Sketch of the random circuit with depth 
$d = 
3$, 
which consists of layers of one-qubit gates and two-qubit gates. CNOT gates 
are 
chosen as two-qubit entangling 
gates. {\bf (d)} Expectation value $\bra{\psi_{{\cal 
R},1}(t)}S_1^z\ket{\psi_{{\cal 
R},1}(t)} \approx 2C_{1,1}(t)$ for random circuits ${\cal R}$ with different 
depths 
$d = 5,10,15,20$. For sufficiently large $d$, data becomes independent of $d$ 
and fluctuations around the exact dynamics of $C_{1,1}(t)$ vanish. Data is 
shown for the one-dimensional Heisenberg chain with system size $L = 25$ and 
Trotter time step $\delta t = 0.5$. Figure adapted from Ref.\ 
\cite{Richter_2021}.}
 \label{Fig::Random}
\end{figure}

We now discuss the individual components of the algorithm 
in more detail. As in \cite{Boixo_2018, Arute_2019}, we consider a 
two-dimensional grid of qubits. The random circuit ${\cal R}$ then consists of 
individual cycles, 
each composed of a layer of one-qubit gates and a layer of two-qubit gates, 
where we denote the total number of cycles with $d$. The two-qubit gates 
generate entanglement between different parts of the system, and different 
choices are possible such as CZ or CNOT gates \cite{Richter_2021, Bensa2021}, or 
other 
hardware-specific gates \cite{Arute_2019, Mi2021}. As shown in Fig.\ 
\ref{Fig::Scheme}~(b), they 
are aligned in one of the patterns {\bf A}-{\bf D}, and the sequence {\bf 
ABCD$\cdots$} is repeated in subsequent cycles throughout ${\cal R}$. Afer $d$ 
cycles, $\ket{\psi_{{\cal R},\ell'}} = \sum_{k =1}^{2^L} c_k \ket{k}$ is a 
superposition of computational basis states. In particular, even for moderately 
shallow ${\cal R}$,   $\ket{\psi_{{\cal R},\ell'}}$ will approximate the 
properties of a full Haar-random state (we will discuss this fact further below 
in the context of Fig.\ \ref{Fig::Random}), i.e., the coefficients $c_k$ are 
approximately distributed according to a Gaussian distribution with zero mean. 
As a consequence, these states can be used within the typicality approach to 
calculate $C_{\ell,\ell'}(t)$. 

As a numerical example, we here consider the 
dynamics with respect to the paradigmatic spin-$1/2$ Heisenberg chain,  
${\cal H} = \sum_{\ell} \left(S_\ell^x S_{\ell+1}^x + S_\ell^y 
S_{\ell+1}^y + S_\ell^z S_{\ell+1}^z\right)$, 
where the one-dimensional model can be readily studied using a snake-like 
path through the two-dimensional grid, see Fig.\ \ref{Fig::Scheme}~(c). The 
time evolution with respect to ${\cal H}$ is then evaluated by decomposing 
$U(t)= \exp(-i{\cal H}t)$ into discrete Trotter steps \cite{DeVries1993, 
Smith2019}, 
\begin{equation}\label{Eq::Trotter}
 U(t) = \left(e^{-i{\cal H}\delta t}\right)^{N} \approx
\left(e^{-i{\cal H}_\text{e}\delta t}e^{-i{\cal H}_\text{o}\delta 
t}\right)^{N} + {\cal O}(\delta t^2)\ , 
\end{equation}
where ${\cal H}_\text{e}$ (${\cal H}_\text{o}$) denotes the even (odd) bonds of 
${\cal H}$ and $\delta t = t/N$ is a (short) time step. \\

The spatiotemporal correlation function $C_{\ell,\ell'}(t)$ can be interpreted 
as a spin excitation moving in front of an infinite-temperature background. As 
shown in Fig.\ \ref{Fig::Scheme}~(c), using $\ell' = 1$, the excitation created 
at the edge spreads through the system under time evolution. In particular, as 
the total magnetization is conserved by ${\cal H}$, i.e., $[{\cal H},\sum_\ell 
S_\ell^z] = 0$, $C_{\ell,\ell'}(t)$ is expected to show hydrodynamic behavior 
at long times \cite{Bertini_2021, Richter2019f}. This hydrodynamic behavior can 
be conveniently analyzed either 
by inspecting the decay of the autocorrelation function $C_{1,1}(t)$ 
or by studying the spatial variance of the full density profile,  
$\Sigma^2(t) = \sum_\ell \ell^2 \widetilde{C}_{\ell,1}(t) - \big[\sum_\ell \ell 
\widetilde{C}_{\ell,1}(t)\big]^2$, 
where $\widetilde{C}_{\ell,1}(t) = C_{\ell,1}(t)/\sum_{\ell=1}^L C_{\ell,1}(t)$ 
with 
$\sum_\ell 
\widetilde{C}_{\ell,1}(t) = 1$. Specifically, their respective power-law 
exponents reflect the type of hydrodynamic transport, $\alpha(t) = -d \ln 
C_{1,1}(t)/d \ln t$, $\beta(t) =d \ln 
\Sigma^2(t)/d \ln t$,    
where for one-dimensional systems $\alpha = 0.5$ corresponds to normal 
diffusion, $\alpha = 1$ indicates ballistic transport, and $0<\alpha < 0.5$  
($0.5<\alpha < 1$) signal anomalous subdiffusion (superdiffusion) 
\cite{Bertini_2021}. \\

After having outlined the general principle of the algorithm, it is insightful 
to go back one step and study the buildup of randomness of $\ket{\psi_{{\cal 
R},\ell'}}$ due to the action of ${\cal R}$. Such an 
analysis is presented in Figs.\ \ref{Fig::Random}~(a) and (b) in terms of the 
participation entropy $S_{p_k}(\ket{\psi_{{\cal  
R},\ell'}})$ and the von Neumann 
entanglement entropy $S_\text{vN}(\ket{\psi_{{\cal R},\ell'}})$, 
\begin{equation}
 S_{p_k}(\ket{\psi_{{\cal 
R},\ell'}}) 
= -
\sum_{k =1}^{2^L} p_k \ln p_k\ ,\qquad S_\text{vN}(\ket{\psi_{{\cal R},\ell'}}) 
= -\text{tr}[\rho_A \ln \rho_A]\ , 
\end{equation}
with $p_k = 
|c_k|^2$, and $\rho_A = \text{Tr}_B\ket{\psi_{{\cal 
R},\ell'}}\bra{\psi_{{\cal R},\ell'}}$ being the reduced density matrix for a 
half-system bipartition. As already mentioned above, 
${\cal R}$ consists of one-qubit and two-qubit gates, cf.\ Fig.\ 
\ref{Fig::Random}~(c). For the simulations shown here, the one-qubit gates are 
drawn at random from the set $\lbrace X^{1/2},Y^{1/2},T\rbrace$, where 
$X^{1/2}$ ($Y^{1/2}$) are $\pi/2$ rotations around the x-axis (y-axis) of the 
Bloch sphere and $T$ is the non-Clifford gate $T = \text{diag}(1,e^{i\pi/4})$. 
Similar to \cite{Arute_2019, Boixo_2018}, gates on a given site have to be 
different in subsequent cycles. As two qubit gates, we here consider CNOT gates. 
As shown in Fig.\ \ref{Fig::Random}~(a) and (b), both $S_{p_k}(\ket{\psi_{{\cal 
R},\ell'}})$ and $S_\text{vN}(\ket{\psi_{{\cal R},\ell'}})$ saturate towards 
their analytically known random-state values already for moderate depth $d$ of 
${\cal R}$. Specifically, $S_{p_k}(\ket{\psi_{{\cal 
R},\ell'}})$ approaches
$\ln(2^{L-1}) - 1 
+ 
\gamma$ with Euler 
constant $\gamma \approx 0.577$ \cite{Boixo_2018} already at $d \lesssim 10$, 
with no 
major dependence on $L$. Likewise, $S_\text{vN}(\ket{\psi_{{\cal R},\ell'}})$ 
approaches the ``Page value'' 
$\ln(2^{L/2})-1$ \cite{Page_1993} appropriate for a random state on $L-1$ sites, 
albeit the depth $d$ to reach this value appears to grow slightly with $L$. 
It is important to stress that the design of ${\cal R}$ chosen here 
is by no means optimized but already sufficient to create highly random and 
entangled states. 
In Fig.\ \ref{Fig::Random}~(d), we analyze the dependence of the expectation value 
$\bra{\psi_{{\cal R},1}(t)}S_1^z\ket{\psi_{{\cal 
R},1}(t)}$ on the depth $d$ of ${\cal R}$. While this expectation 
value converges to the exact autocorrelation function $2C_{\ell,\ell'}(t)$ for 
sufficiently large $d$, we find in Fig.\ \ref{Fig::Random}~(d) that even for 
rather shallow circuits with $d = 10$, the resulting dynamics is almost 
indistinguishable from the dynamics for $d = 20$. Thus, even for states with 
entanglement below the Page value, cf.\ Fig.\ \ref{Fig::Random}~(b), the 
resulting dynamics is still a good approximation to $C_{1,1}(t)$ and captures 
the correct hydrodynamic behavior. \\

\begin{figure}[tb]
 \centering
 \includegraphics[width = 0.8\textwidth]{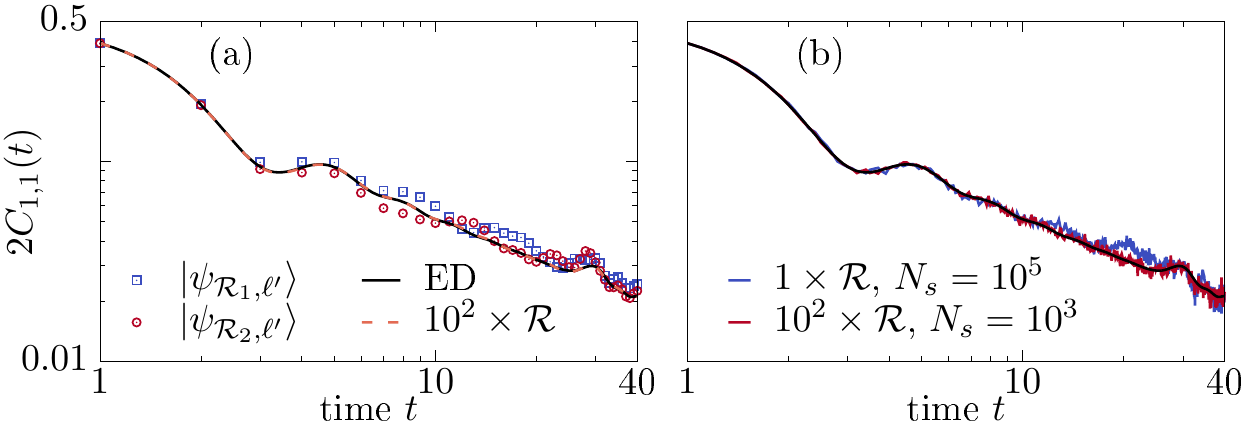}
 \caption{Accuracy of the typicality approach in comparison with exact 
diagonalization for a one-dimensional Heisenberg chain with $L=16$. {\bf (a)} 
The 
full expectation value $\bra{\psi_{{\cal R},\ell'}(t)} S_\ell^z 
\ket{\psi_{{\cal R},\ell'}(t)}$ for two different individual realizations of 
the random circuit (${\cal R}_1$, ${\cal R}_2$) shows visible fluctuations, 
while averaging the expectation value over $100$ random instances of ${\cal R}$ 
yields results indistinguishable from ED. {\bf (b)} Reconstructed expectation 
value 
from sampling the output of the circuit $N_s$ times as required in an actual 
experiment, cf.\ Eq.\ \eqref{Eq::Reconstruct}. Note that $N_s$ here denotes the 
number of samples for each individual realization of ${\cal R}$. In particular, 
averaging over ${\cal R}$ and sampling over $N_s$ can be combined with each 
other. Figure adapted from Ref.\ \cite{Richter_2021}.}
 \label{Fig::Accuracy}
\end{figure}

Finally, let us comment on the accuracy of the typicality approach. In Fig.\ 
\ref{Fig::Accuracy}~(a), we compare the dynamics of two states that 
result from different realizations of the random circuit, ${\cal R}_1$ and 
${\cal R}_2$. Moreover, we also show data obtained from exact diagonalization. 
Even for the rather small system size $L = 16$ chosen here, we find that the 
dynamics obtaind from $\ket{\psi_{{\cal R}_1,\ell'}}$ and  $\ket{\psi_{{\cal 
R}_2,\ell'}}$ closely follow the exact result with only minor fluctuations. In 
particular, these residual fluctuations can be further suppressed by averaging 
over multiple random realizations of ${\cal R}$. As shown in Fig.\ 
\ref{Fig::Random}~(a), the dynamics obtained by averaging over $10^2$ 
realizations is indistinguishable from ED. Note that for larger system sizes 
(i.e., significantly larger Hilbert-space dimensions) quantum typicality 
becomes more and more accurate such that such an averaging is not necessary 
anymore and a single realization of ${\cal R}$ is sufficient \cite{Jin_2021}. \\

While we have so far focused on the full expectation value 
$\bra{\psi_{{\cal R},\ell'}(t)} S_\ell^z 
\ket{\psi_{{\cal R},\ell'}(t)}$, this expectation value cannot be obtained on a 
quantum computer in a single run. In particular, $\ket{\psi_{{\cal 
R},\ell'}(t)}$ will generally be a superposition of computational basis states, 
$\ket{\psi_{{\cal R},\ell'}(t)} = \sum_{k=1}^{2^L} a_k \ket{k}$, 
and the coefficients $a_k$ can be reconstructed by repeating the experiment 
multiple times and averaging the final measurements of the qubits. More 
specifically, $C_{\ell,\ell'}(t)$ can be reconstructed as, 
\begin{equation}\label{Eq::Reconstruct} 
2C_{\ell,\ell'}(t) = \frac{1}{2}\left(\sum_{\ket{k}, \ell = \uparrow} 
|\widetilde{a}_k|^2 - \sum_{\ket{k}, \ell = \downarrow} |\widetilde{a}_k|^2 
\right)\ , 
\end{equation}
where $|\widetilde{a}_k|^2$ is the experimentally obtained probability of the 
state $\ket{k}$, and the two sums in \eqref{Eq::Reconstruct} run over all 
states for which the spin $\ell$ is up or down respectively. The accuracy can 
be systematically improved by increasing the number of samples $N_s$ such that 
 $|\widetilde{a}_k|^2 \to |a_k|^2$. 
Crucially, as we show in Fig.\ \ref{Fig::Accuracy}~(b), the sampling of the 
distribution of the $|a_k|^2$ can be combined with the averaging over 
random realizations of ${\cal R}$. Specifically, Fig.\ \ref{Fig::Accuracy}~(b) 
compares the dynamics obtained from one realization of ${\cal R}$ with $N_s = 
10^5$ repetitions to the dynamics obtained from $10^2$ realizations of ${\cal 
R}$ with $N_s = 10^3$ repetitions each, i.e., the total number of runs is the 
same in both cases. While the noise level is found to be similar in both cases, 
we observe that the data averaged over multiple ${\cal R}$ agrees better with 
ED. In this context, we note that varying ${\cal R}$ on a NISQ device is 
straightforward experimentally.

\colorlet{shadecolor}{blue!80}
\begin{shaded}
\noindent 
 \textcolor{white}{\bf Box 2 $|$ QUANTUM TYPICALITY}
\end{shaded}
\vspace{-2mm}
\colorlet{shadecolor}{blue!10}
\begin{shaded}
\noindent
The notion of quantum typicality refers to the fact that the overwhelming 
majority (Haar measure) of quantum states within some energy shell yields 
expectation values of observables very close to the full microcanonical 
ensemble \cite{Lloyd1988, Goldstein2006, Popescu2006, Reimann2007}. Speaking 
differently, given a single pure quantum 
state, drawn at random from a high-dimensional Hilbert space, the 
expectation values of observables with respect to this state will be 
effectively indistinguishable from their thermodynamic equilibrium values. As 
such, quantum typicality has been put forward as an important concept to explain 
the emergence of thermodynamic behavior in closed quantum systems 
\cite{Gemmer_2010, Balz_2018}. Here, we focus on the implications of typicality 
for efficient simulations (see \cite{Jin_2021, Heitmann_2020} for reviews).

\noindent
{\bf Quantum typicality as a numerical tool} 

\noindent
The key idea in this context is that random quantum states are highly accurate 
{\it trace estimators} \cite{Hams2000, Schnack_2020}. To demonstrate this 
fact, let $\ket{\psi}$ be a 
random state, drawn from the unitarily invariant Haar measure, 
\begin{equation}
 \ket{\psi} = \sum_{k = 1}^D (a_k + ib_k)\ket{k}\ , 
\end{equation}
where the coefficients $a_k$ and $b_k$ are Gaussian-distributed 
with zero mean and unit variance, and $\ket{k}$ denote a set of orthogonal 
basis states of the Hilbert space with dimension $D$. Given an observable ${\cal 
O}$, its trace can then be written as~\cite{Jin_2021}, 
\begin{equation}\label{Eq::Trace}
 \text{tr}[{\cal O}] =  \frac{D\overline{\bra{\psi} {\cal O} 
\ket{\psi}}}{\overline{\braket{\psi|\psi}}}\ ,\qquad \text{tr}[{\cal O}] 
\approx  \frac{D\bra{\psi} {\cal O} 
\ket{\psi}}{\braket{\psi|\psi}}\ , 
\end{equation}
where the overline means averaging over random realizations 
of $\ket{\psi}$. Importantly, the second equation in 
\eqref{Eq::Trace} emphasizes that for sufficiently large Hilbert-space 
dimensions, even a single realization of $\ket{\psi}$ provides an accurate 
approximation of $\text{tr}[{\cal O}]$. In particular, given some mild 
assumptions on the spectral properties of ${\cal O}$, the statistical error of 
the approximation vanishes for $D\to\infty$ (see, e.g., \cite{Jin_2021} for 
details on error bounds). It is straightforward to extend the scheme in Eq.\ 
\eqref{Eq::Trace} to equilibrium expectation values of ${\cal O}$ at 
temperature $T = 1/\beta$ \cite{Sugiura2012},   
\begin{equation}\label{Eq::Thermal}
 \langle {\cal O} \rangle_\text{eq} = \frac{\text{tr}[{\cal O}e^{-\beta 
{\cal H}}]}{\text{tr}[e^{-\beta {\cal H}}]} = \frac{\bra{\psi_\beta} {\cal 
O}\ket{\psi_\beta}}{\braket{\psi_\beta|\psi_\beta}} + \epsilon\ ,\quad 
\ket{\psi_\beta} = e^{-\beta {\cal H}/2}\ket{\psi}\ , 
\end{equation}
where $\ket{\psi_\beta}$ is sometimes referred to as thermal pure quantum 
state \cite{Sugiura2012} and the standard deviation of the statistical error  
scales as $\sigma(\epsilon) \propto 1/\sqrt{d_\text{eff}}$, with $d_\text{eff} 
= \text{tr}[\exp(-\beta({\cal H}-E_0))]$ being an effective dimension, and 
$E_0$ is the ground-state energy of ${\cal H}$ \cite{Jin_2021, Sugiura2012}. 
Furthermore, quantum 
typicality also carries over to time-dependent situations (see e.g., 
\cite{Bartsch2009, Balz_2018, Monnai2014, Richter2019a,Endo2018}), which has 
been particularly exploited in the context of dynamical 
two-point correlation functions, which can be approximated 
as~\cite{Elsayed2013, Iitaka2004, Steinigeweg2014, Richter2019b},  
\begin{equation}\label{Eq::2PCorr}
 \langle {\cal O}_1(t) {\cal O}_2\rangle_\text{eq} = 
\frac{\text{tr}[{\cal O}_1(t){\cal O}_2e^{-\beta {\cal 
H}}]}{\text{tr}[e^{-\beta {\cal H}}]} \approx \frac{\bra{\psi_\beta(t)}{\cal 
O}_1\ket{\varphi_\beta(t)}}{\braket{\psi_\beta(0)|\psi_\beta(0)}}\ , 
\end{equation}
where $\ket{\varphi_\beta(t)} = e^{-i{\cal H}t} {\cal O}_2 e^{-\beta {\cal 
H}/2}\ket{\psi}$, $\ket{\psi_\beta(t)} = e^{-i{\cal H}t} e^{-\beta {\cal 
H}/2}\ket{\psi}$. The crucial ingredient why the pure-state approximations in 
Eqs.\ \eqref{Eq::Thermal} and \eqref{Eq::2PCorr} are useful is given by the 
fact that the temperature- and time-dependence can be efficiently evaluated by 
using well-established sparse-matrix techniques \cite{Fehske_2009} (e.g., 
Krylov-subspace methods \cite{Nauts1983}, 
or Chebyshev-polynomial expansions \cite{TalEzer1984, Dobrovitski2003, 
Richter2020_2}). These methods essentially rely on 
matrix-vector multiplications which can be carried out both time and memory 
efficient. In particular, no exact diagonalization of ${\cal H}$ is required 
such that significantly larger Hilbert spaces can be handled. 

\noindent
{\bf Quantum typicality versus eigenstate thermalization hypothesis (ETH)} 

\noindent
The ETH asserts that the diagonal matrix elements of observables written in 
the eigenbasis of chaotic nonintegrable Hamiltonians are a smooth function of 
energy and agree with the corresponding microcanonical expectation 
value~\cite{Dalessio2016}, 
\begin{equation}
 {\cal O}_{nn} = \bra{n}{\cal O}\ket{n} \approx {\cal O}_\text{mc}(E)\ ,\qquad 
{\cal O}_\text{mc}(E) = \frac{1}{N_E}\sum_{E_n \approx E} \bra{n}{\cal 
O}\ket{n}\ . 
\end{equation}
While the ETH has been numerically confirmed for a variety of models and 
observables, it is known to break down for instance in strongly disordered 
models exhibiting many-body localization \cite{Abanin_MBL}. In 
this case, the 
diagonal matrix elements show pronounced eigenstate-to-eigenstate fluctuations, 
${\cal O}_{nn} \neq {\cal O}_{(n+1)(n+1)} \neq {\cal O}_\text{mc}(E)$. In 
contrast, the concept of typicality remains valid irrespective of the ETH being 
fulfilled. Specifically, consider a state $\ket{\psi_E}$ that is constructed as 
a random superposition of eigenstates $\ket{n}$ in an energy shell with mean 
energy $E$. Then, the expectation value of ${\cal O}$ with respect to 
$\ket{\psi_E}$ will again be very close to the microcanonical expectation value 
as long as the energy window contains sufficiently many eigenstates to reduce
statistical errors,
\begin{equation}
 \frac{\bra{\psi_E} {\cal O} \ket{\psi_E}}{\braket{\psi_E|\psi_E}} \approx {\cal O}_\text{mc}(E)\ ,\qquad    
\ket{\psi_E} = \sum_{E_n \approx E} c_n \ket{n}\ ,\ \text{where $c_n$ are 
Gaussian random numbers.}
\end{equation}
As a consequence, quantum typicality can be exploited as a numerical tool to 
study the properties of integrable or many-body localized systems where the ETH 
breaks down \cite{Richter_2020, Richter_2018, Steinigeweg_2016, Luitz_2017}, as 
well as to verify the validity of the ETH for system sizes beyond the range of 
exact diagonalization \cite{Steinigeweg2014a, Wang2021}.  

\end{shaded}

\section{Conclusion}

The goal of this chapter was to give a brief overview of certain aspects of 
random quantum circuits. Such random circuits have recently gained increased attention 
as minimally structured models for quantum many-body systems, providing new 
insights into challenging questions, e.g., regarding the dynamics of entanglement or the 
emergence of hydrodynamics. In view of the vast literature, we here deliberately refrained 
from giving an in-depth review of all actively pursued directions, but rather refer the interested reader to 
the pertinent references included in Secs.\ \ref{Sec::Intro} - \ref{Sec::NISQ}. Instead, we here 
focused in more detail on two particular topics, i.e., (i) entanglement transitions in 
monitored circuits, where random unitary gates are interspersed with measurements, and (ii) random quantum circuits recently implemented 
on noisy intermediate-scale quantum devices to achieve a quantum computational advantage. 

In the context of monitored circuits, we have given an introduction to the notion of quantum trajectories and reviewed 
the equivalent pictures of a measurement-induced pure-state transition from volume-law to area-law scaling of the 
entanglement entropy and a purification transition from mixed to pure states with increasing measurement rate. We have particularly discussed the 
critical properties of the transition in one and two dimensions and explained how Clifford circuits, especially in combination with graph states, provide a powerful approach to explore this type of physics 
numerically. Moreover, we have touched on the challenges to realize and observe entanglement transitions in monitored circuits in actual experiments. 
While our understanding of measurement-induced criticality has increased substantially over the last couple of years, a variety of open questions still 
remain. For example, the exact nature of the critical point in realistic limits remains open, despite recent progress~\cite{nahumMeasurementEntanglementPhase2021}. It has also been shown that it is possible to stabilize phases of matter using specific measurement protocols~\cite{lavasaniTopologicalOrderCriticality2020,lavasaniMeasurementinducedTopologicalEntanglement2021,sangMeasurementProtectedQuantum2020}, which would otherwise be unstable in thermal equilibrium. However, it is not yet clear what restrictions, if any, apply to the phases realizable in this way. Similar protection against equilibration can be afforded by many-body localization~\cite{huseLocalizationprotectedQuantumOrder2013}, but this is not possible in the presence of non-Abelian symmetries~\cite{potterSymmetryConstraintsManybody2016}, for example, so it would be interesting if monitored quantum circuits could provide a route to avoiding these restrictions. Finally, it would be interesting to see if the quantum error-correcting properties of the volume-law phase, which have provided a useful lens on the phenomenology~\cite{choiQuantumErrorCorrection2020,gullansDynamicalPurificationPhase2020,fanSelfOrganizedErrorCorrection2020,liStatisticalMechanicsQuantum2020,liRobustDecodingMonitored2021}, could be put to practical use in quantum computers. While the hybrid dynamics encode quantum information, and a decoder is known to exist in the volume-law phase~\cite{gullansDynamicalPurificationPhase2020}, it is not clear whether in general a `good' decoder exists which could efficiently detect and correct errors~\cite{hastingsDynamicallyGeneratedLogical2021a}.

Regarding random circuits on NISQ devices, we have provided a brief introduction to the 
ideas of random-circuit sampling and cross-entropy benchmarking to achieve a quantum computational advantage.
Furthermore, we have focused on a recently proposed random-circuit based algorithm to simulate 
hydrodynamics on NISQ devices \cite{Richter_2021}, which emphasizes that random circuits are not 
just abstract tools to outperform classical computers for a specific computational problem, but are relevant 
also for a wider range of applications. In this context, we have demonstrated that random circuits can efficiently 
generate random and highly entangled quantum states. Importantly, we have explained that such random states are 
immediately useful for simulations, both on quantum and classical computers, by leveraging the concept of of quantum typicality.
Exploring in more detail the potential applications of random quantum states and quantum typicality on NISQ 
devices will be an interesting direction of future research. Such applications are particularly appealing from a complexity point 
of view as random states can be prepared on NISQ devices by random circuits of moderate depth. However, due to their high entanglement, 
they are usually not amenable to concepts such as matrix-product states, which 
makes classical representations very costly. Applications of random 
states thus fall naturally into a category that is classically hard, but might be 
accessible by NISQ devices. Let us note that quantum 
typicality has in fact been recently exploited in experiments on actual quantum 
hardware. In \cite{Mi2021}, a scheme very similar to that outlined in Sec.\ 
\ref{Sec::Hydro_on_QC} 
was used to calculate dynamical spin-spin correlation functions in a driven 
Floquet spin chain in order to verify the occurrence of discrete 
time-crystalline eigenstate order. Moreover, in \cite{Powers2021}, quantum 
typicality was 
used to evaluate thermodynamic expectation values at finite temperature (cf.\ 
Box 2), which involved the approximation of the imaginary 
time-evolution of random states on a NISQ device \cite{Motta2019}.   

\section*{Acknowledgements}
We acknowledge discussions with Sougato Bose, Matthew Fisher, Andrew Green, Michael Gullans, David Huse, Lluis Masanes, Jed Pixley, Robin Steinigeweg, Marcin Szyniszewski, Curt von Keyserlingk, and Brayden Ware. 
This work was funded by the 
European Research 
Council 
(ERC) under the European Union's Horizon 2020 research and innovation programme
(Grant agreement No.\ 853368). 

\bibliographystyle{apsrev4-2}

\bibliography{Entanglement_Chapter}

\end{document}